\documentclass[aps,prl,reprint,superscriptaddress,showpacs]{revtex4-1}

\usepackage{graphicx}
\usepackage{upgreek}
\usepackage{multirow}

\usepackage{psfrag}
\usepackage{epsfig}
\usepackage{epstopdf} 

\usepackage{amssymb}
\usepackage{amsmath}
\usepackage{xspace}
\usepackage{bbold}

\begin{document}


\title{Observation of infinite-range intensity correlations above, at and below\\ the mobility edges of the 3D Anderson localization transition}


\author{W.\,K. Hildebrand}
\author{A. Strybulevych}
\affiliation{Department of Physics and Astronomy, University of Manitoba, Winnipeg, Canada}

\author{S.\,E. Skipetrov}
\author{B.\,A. van Tiggelen}
\affiliation{Universit\'{e} Grenoble 1/CNRS, LPMMC UMR 5493, B.P. 166, 38042 Grenoble, France}
\author{J.\,H. Page}
\email[]{John.Page@umanitoba.ca}
\affiliation{Department of Physics and Astronomy, University of Manitoba, Winnipeg, Canada}


\date{\today}

\begin{abstract}
We investigate long-range intensity correlations on both sides of
the Anderson transition of classical waves in a three-dimensional (3D) disordered material. Our ultrasonic experiments are designed to unambiguously detect a recently predicted infinite-range $C_0$ contribution, due to local density of states fluctuations near the source.  We find that these $C_0$ correlations, in addition to $C_2$ and $C_3$ contributions, are significantly enhanced near mobility edges. Separate measurements of the inverse participation ratio reveal a link between $C_0$ and the anomalous dimension $\Delta_2$, implying that $C_0$ may also be used to explore the critical regime of the Anderson transition.
\end{abstract}

\pacs{42.25.Dd, 43.20.Gp, 71.23.An, 64.60.al}

\maketitle


The phenomenon of Anderson localization---the halt of wave transport due to destructive interferences of scattered waves---was first discovered for electrons in disordered solids \cite{Anderson1958,Abrahams1979a,Vollhardt1980b,Evers2008a,Sheng2006a,Abrahams2010}. John \cite{John1983,John1984} and Anderson \cite{Anderson1985} later suggested that it may also take place for classical waves, such as sound or light. The latter open up new ways to study Anderson localization that would be difficult, or even impossible, to implement in electronic systems.
Time- and position-resolved measurements, for example, have enabled the first unambiguous observation of three-dimensional (3D) Anderson localization of elastic waves \cite{Hu2008b} and yield promising results for light \cite{Sperling2013}.
Further insight into this unique regime of wave physics can be gained by investigating the correlations of the intensity fluctuations that constitute speckle patterns.
While short- and long-range correlations of the intensity (denoted ${C_1, C_2,C_3}$) 
\cite{Feng1988,DeBoer1992,Berkovits1994,Scheffold1997a,VanRossum1999a,Sebbah2000,Chabanov2004,Akkermans2007},
and even phase \cite{Genack1999,Cowan2007a}, have been predicted and observed in the regime of weak disorder, they remain unexplored in the localized regime and at the mobility edge (ME) where the transition between diffuse and localized behavior occurs.
Moreover, a new type of infinite-range intensity correlation (denoted $C_0$), originating from scattering in the vicinity of the source, has recently been predicted \cite{Shapiro1999a, Skipetrov2000}.
For a point source embedded in a disordered medium, this correlation was shown to be closely related to fluctuations of the local density of states (LDOS) at the source position \cite{VanTiggelen2006b,Caze2010}.
Hence, the recent measurements of LDOS fluctuations
\cite{Birowosuto2010a,Krachmalnicoff2010a,Sapienza2011,Garcia2012} can be considered as indirect evidence for $C_0$ even though some caution may be required depending on the source type \cite{El-Dardiry2011}.
LDOS fluctuations are expected to grow as the states become spatially localized \cite{Krachmalnicoff2010a,Garcia2012,Mirlin2000a},
with recent theoretical studies even reporting their variance to behave as a one-parameter scaling function of sample size and localization length \cite{Dobrosavljevic2003a,Murphy2011}, which means they constitute a new tool to provide insight into the Anderson transition.
In view of the profusion of results concerning LDOS fluctuations, it is remarkable that no \emph{direct} measurement of the $C_0$ contribution to the intensity correlation function has been reported so far \cite{Cinf}.

In this Letter, we present the first direct experimental evidence of infinite-range ($C_0$) spatial and frequency correlations of intensity above, at and below the ME of the Anderson transition of a disordered, strongly scattering 3D material.
The experiments were performed using ultrasonic techniques on samples in which 3D Anderson localization of ultrasound has been demonstrated previously \cite{Hu2008b}. Comparison of experiment with theory, coupled with complementary measurements designed to suppress infinite-range correlations when desired, allows the $C_0$ contribution to the correlations to be clearly separated from the other contributions
($C_1$, $C_2$, or $C_3$), unambiguously revealing the presence of large infinite-range correlations. We observe that these correlations grow dramatically near the ME in our samples.
Motivated by the prediction that the LDOS fluctuations are closely related to multifractality of the wave functions through the $q=2$ generalized inverse participation ratio (gIPR) \cite{Krachmalnicoff2010a,Murphy2011},
we measure the anomalous dimension $\Delta_2$ for our samples in independent experiments and find good correspondence between this quantity and measured $C_0$ correlations.  This clearly demonstrates the link between multifractality, $C_0$, and the LDOS fluctuations.

The samples investigated are disordered elastic networks of aluminum beads, weakly brazed together to form slabs 
(see supplemental material \cite{suppmat}).
This porous mesoscale structure leads to very strong scattering with low absorption in the frequency range investigated ($\sim0.5\text{--}2.5$~MHz), a crucial feature for the observation of 3D Anderson localization of ultrasound in this material. The mesoscale structure also leads to high contrast in the density of states of the aluminum matrix compared to that of the pores---yet another reason for anticipating strong fluctuations of the LDOS. The samples were waterproofed so that the experiments could be performed in a water tank with either vacuum or air in the pores, thereby ensuring that the detected transmitted waves had traveled only through the aluminum bead network.

In our experiments, a tightly focused broadband ultrasonic pulse (with beam waist smaller than the
wavelength) is incident on the sample, and the transmitted pressure is detected in the near field by a sub-wavelength hydrophone \cite{suppmat}.
To capture contributions to $C_0$ due to LDOS fluctuations at both the focal point of the incident wave (source point) and the detector, we scan both the source and detector over the surface of the sample. The recorded pressure fields
$p(\mathbf{r}, t)$ are Fourier transformed to obtain the intensity $I(\mathbf{r}, \omega) \propto |p(\mathbf{r}, \omega)|^2$ as a function of frequency for each pair of source and detector positions.  The intensity correlation function is calculated as
\begin{equation}
C_{\omega}(\Delta r, \Omega) =
\frac{\langle \delta I(\mathbf{r}, \omega-\frac12 \Omega)
\delta I(\mathbf{r} + \Delta \mathbf{r}, \omega+\frac12 \Omega) \rangle}{\langle I(\mathbf{r}, \omega-\frac12 \Omega) \rangle \langle I(\mathbf{r}+\Delta \mathbf{r}, \omega+\frac12 \Omega) \rangle},
\end{equation}
where the angular brackets denote ensemble averaging and
$\delta I = I - \langle I \rangle$ is the fluctuation of the intensity. Ensemble averaging is done by scanning over many source and detector positions corresponding to the same $\Delta r$. For comparison, experiments with a single (stationary) source point were also performed, in which case the ensemble averaging was done only over all possible detector positions; this suppresses $C_0$ correlations due to LDOS fluctuations at the source.
In what follows we will study spatial correlations $C_{\omega}(\Delta r) = C_{\omega}(\Delta r, 0)$ and frequency correlations $C_{\omega}(\Omega) = C_{\omega}(0, \Omega)$ separately.

\begin{figure}
    \includegraphics[width=0.95\columnwidth]{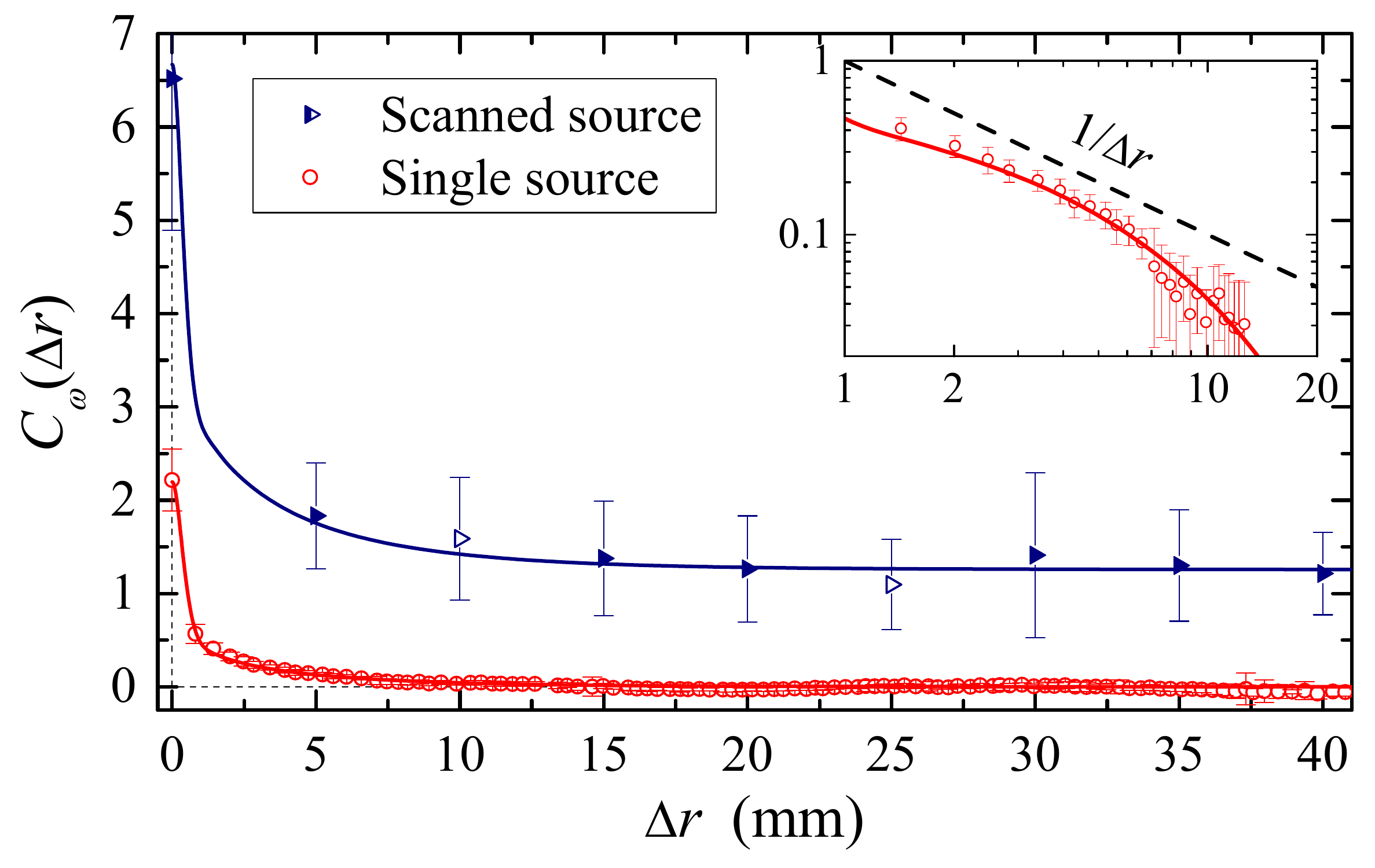}
	\caption{\label{fig:spatcorr1} (color online) Spatial intensity correlations for the two types of experiments at 2.4~MHz. The scanned-source data show convincing evidence of infinite-range ($C_0$) correlations, which are suppressed when only a single source point is used. Lines are theoretical fits using the values of parameters given in Table \ref{tab:fit}.
The inset shows the single-source data on a log-log scale in order to reveal the extent to which the expected $1/\Delta r$ dependence is observed for $C_2$ and $C_3$ at intermediate length scales. Data have been averaged over a bandwidth of 750~kHz, and the error bars are the standard deviations associated with the data's statistical fluctuations, which are observed to be inherently large near the Anderson transition. (The open symbols for the scanned-source data represent positions where the measurements are not as reliable because of a smaller signal-to-noise ratio.)}
	\end{figure}

Figure~\ref{fig:spatcorr1}
shows the spatial correlations measured near $f = \omega/2\pi = 2.4$ MHz, the frequency at which Anderson localization of elastic waves was demonstrated in this sample \cite{Hu2008b}.  For both types of experiments, the correlations decay rapidly at small
$\Delta r$ due to $C_1$, with a slower decay due to $C_2$ and $C_3$ that extends out to
$\Delta r \sim 10$~mm, beyond which $C_{\omega}(\Delta r)$ becomes independent of distance.  For the data where the source position is varied, an asymptotic value of order unity is seen for the correlations, showing clear evidence of a $C_0$ term due to LDOS fluctuations at the source.
By contrast, no infinite-range correlations are seen for the single-source data, consistent with the fact that the LDOS at the source position does not fluctuate in this case.

To gain further insight into this behavior, we compare our experimental data with theoretical calculations.
We compute $C_1$, $C_2$, $C_3$, and $C_0$ correlation functions assuming weak disorder ($k\ell\gg1$, where $k$ is the wavenumber in the medium and $\ell$ is the mean free path) and write the full correlation $C_{\omega}(\Delta r)$ as a function of three fit parameters: $A$, $C_0^{\text{(in)}}$, and $C_0^{\text{(out)}}$.
Although this calculation is not exact, the parametrization into four fundamentally different classes of speckle correlations involving phenomenological constants should be valid even in the critical regime.
The parameter $A$ quantifies the magnitude of $C_2$ and $C_3$ correlations,
$C_0^{\text{(in)}}$ characterizes the magnitude of the genuine $C_0$ correlation due to the LDOS fluctuations at the source point, and
$C_0^{\text{(out)}}$ measures the amplitude of the short-range contribution to $C_0$ due to scattering in the vicinity of both detectors when the latter are close to each other \cite{suppmat}.
$C_0^{\text{(in)}}$ is the asymptotic value of $C_{\omega}(\Delta r)$ for $\Delta r \to \infty$.
The solid lines in Fig.~\ref{fig:spatcorr1} show the results of performing a \emph{joint} weighted fit of these theoretical predictions to both the single- and scanned-source data, thereby determining the values of the parameters shown in Table~\ref{tab:fit}.
In this fit, we account for the fact that $C_0^{\text{(in)}}$ contributes only to the scanned-source correlations and set $C_0^{\text{(in)}} = 0$ for fitting the single-source data; also, since the detector geometry is the same for both experiments, $C_0^{\text{(out)}}$ is constrained to have a common value for the two curves.
Note that for white-noise uncorrelated disorder and point-like source and detector in an infinite disordered medium, $C_0^{\text{(in)}} = C_0^{\text{(out)}} = \pi/k\ell$  \cite{Shapiro1999a}.
In our experiments, however, both the source and detector have finite extent (which differs in each case \cite{suppmat}), and the finite size of the aluminum beads inevitably results in some short-range structural correlations. Therefore, we expect in general that
$C_0^{\text{(in)}} \neq C_0^{\text{(out)}} \ne \pi/k\ell$ \cite{Skipetrov2000}.
Figure~\ref{fig:spatcorr1} provides strong evidence that the large asymptotic value of $C_{\omega}(\Delta r \rightarrow \infty) = C_0^{\text{(in)}} \sim 1$ for the scanned-source experiment is due to $C_0$ correlations.

\begin{table}
\caption{\label{tab:fit} Fit parameters, with uncertainties in parentheses.
The uncertainties are given by the standard deviation of the parameters.
For a point source and detector, the normalized variance, $C(0,0)$, depends on all three parameters: $C(0,0)=1+2[A + C_0^{\text{(in)}} + C_0^{\text{(out)}}]$.
By contrast, the infinite range contributions depend independently on the different contributions to $C_0$, with the asymptotic values of the scanned-source $C(\Delta r,0)$, the single-source $C(0,\Omega)$, and the scanned-source $C(0,\Omega)$ being equal to $C_0^{\text{(in)}}$, $C_0^{\text{(out)}}$, and $C_0^{\text{(in)}}+C_0^{\text{(out)}}$, respectively. }
\centering
\begin{ruledtabular}
\begin{tabular}{lcccc}
Parameter&~2.4 MHz~&~0.97 MHz~&~1.07 MHz~&~1.11 MHz~\\ [0.3ex]
\colrule
\multicolumn{5}{c}{Spatial correlations}\\ [0.3ex]
\colrule
A (single)& 0.50\,(0.02) & 0.48\,(0.02) & 1.29\,(0.04) & 1.59\,(0.06) \\[0.3ex]
A (scanned) & 2\,(1) & 0.8\,(0.2) & 6\,(1) & 6\,(3) \\[0.3ex]
$C_0^{\text{(in)}}$ & 1.3\,(0.2) & 0.42\,(0.02) & 1.06\,(0.08) & 7.8\,(0.5) \\[0.3ex]
$C_0^{\text{(out)}}$ & 0.4\,(0.2) & 0.8\,(0.2) & 1.5\,(0.4) & 7\,(1) \\[0.3ex]
\colrule
\multicolumn{5}{c}{Frequency correlations}\\ [0.3ex]
\colrule
A (single)&  & 0.2\,(0.2) & 0.8\,(0.3) &  \\[0.3ex]
A (scanned) & & 0.7\,(0.1) & 5.2\,(0.8) & \\[0.3ex]
$C_0^{\text{(in)}}$ & & 0.32\,(0.03) & 0.9\,(0.3) &  \\[0.3ex]
$C_0^{\text{(out)}}$  & & 0.62\,(0.06) & 1.3\,(0.1) & \\[0.3ex]
$\Omega_{\text{Th}}/2\pi~\text{(kHz)}$ & & 4.35\,(0.09) & 7.2\,(0.1) & \\
\end{tabular}
\end{ruledtabular}

\end{table}

\begin{figure}
    \includegraphics*[width=8.5cm]{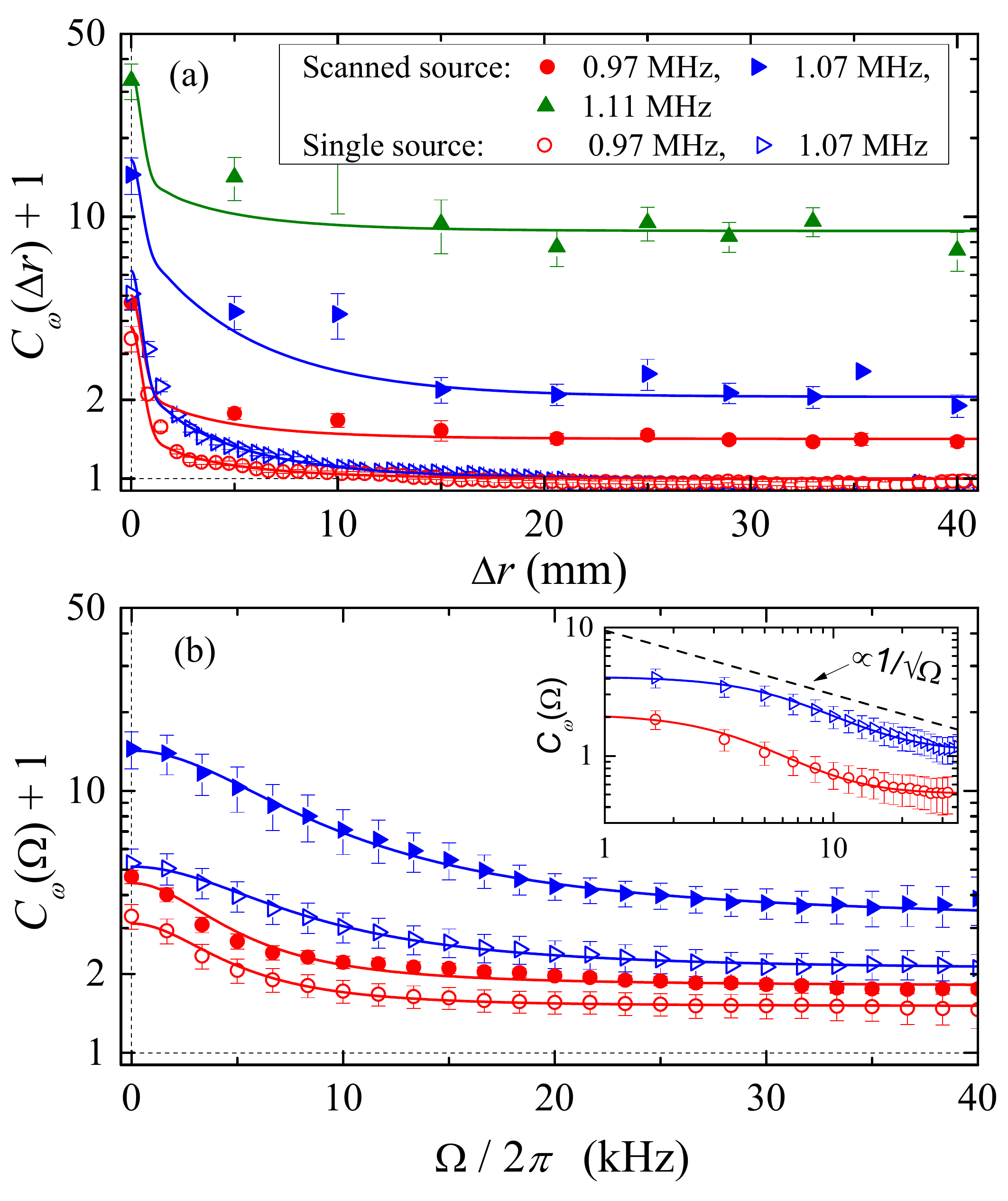}
	\caption{\label{fig:sfcorr} (color online) Spatial (a) and frequency (b) correlations measured near 1~MHz, showing the increase of long-range correlations near a mobility edge. The inset shows the characteristic $1/\sqrt{\Omega}$ behavior expected for $C_2$ and $C_3$ correlations.
For both plots, data have been averaged over a bandwidth of 25~kHz, except at the highest frequency (1.11~MHz), where the data are changing too rapidly with frequency to be meaningfully averaged.  These rapid variations with frequency near 1.11~MHz also complicate measurements of frequency correlations, which are therefore not shown here. The error bars are calculated as in Fig.~\ref{fig:spatcorr1}. Lines show the fits using the parameters given in Table~\ref{tab:fit}.}
	\end{figure}

Similar behavior, with $C_0^{\text{(in)}}\sim 1$, is observed over a broad frequency range from 1.6 to 2.8~MHz, where independent measurements of the dynamic transverse confinement of the transmitted intensity \cite{Hu2008b} indicate that ultrasound is still localized, with similar values of the localization length
$\xi$ ($\xi \approx L = 14.5$~mm for this sample in this frequency range).  At lower frequencies, at least one ME must exist, since previous measurements on these samples revealed diffusive behavior at the much lower frequency of 200~kHz \cite{Hu2008b}.
To investigate the long-range correlations as a ME is approached, experiments were performed at intermediate frequencies, between these well established diffusive and localized regimes \cite{Hu2008b}, with representative data near 1~MHz being presented in Fig.~\ref{fig:sfcorr}.
Both spatial and frequency correlations increase significantly with frequency when a ME, which we estimate to be at approximately 1.1~MHz, is approached. In particular, the asymptotic value of the scanned-source spatial correlations,
$C_0^{\text{(in)}}$, increases from 0.4 to almost 8 over the range of frequencies illustrated in Fig.~\ref{fig:sfcorr}(a).

The frequency correlations also show large increases in $C_0$ over this frequency range [see Fig.~\ref{fig:sfcorr}(b)] \cite{freqfit}.
$C_{\omega}(\Omega)$ contains infinite-range contributions from scattering both near the source and near the detector,
i.e. both $C_0^{\text{(in)}}$ and $C_0^{\text{(out)}}$ contribute to the asymptotic value of
$C_{\omega}(\Omega)$ for large $\Omega$ \cite{suppmat}.
The single-source measurements (which suppress $C_0^{\text{(in)}}$) show that $C_0^{\text{(out)}}$ increases from $0.6$ to $1.3$ between 0.97 and 1.07~MHz.
By comparing the best-fit values (see Table~\ref{tab:fit}), we see that for the scanned-source case, $C_0^{\text{(in)}}$ and $C_0^{\text{(out)}}$ are of the same order of magnitude, as could be expected from the roughly symmetric arrangement of the experiment \cite{C0robust}.

The $C_2$ and $C_3$ correlations, quantified by the parameter $A$, also increase with frequency around 1 MHz, as found from the data for both spatial and frequency correlations (see Fig.~\ref{fig:sfcorr} and Table~\ref{tab:fit}). Because $A \propto 1/(k \ell^*)^2$ to leading order \cite{suppmat}, where $\ell^*$ is the transport mean free path, the increase of $A$ corresponds to a decrease of $k \ell^*$ as the ME is approached. In addition, the values of $A$ found from the fits are always larger in the scanned-source case. Keeping the source fixed not only suppresses $C_0^{\text{(in)}}$ but also reduces the magnitude of long-range $C_2$ and $C_3$ correlations because the latter contain contributions from scattering in the vicinity of the source. This effect does not preclude the clear identification of $C_0$ that stands out by its infinite range in both space and frequency.

\begin{figure}
\includegraphics[width=0.95\columnwidth]{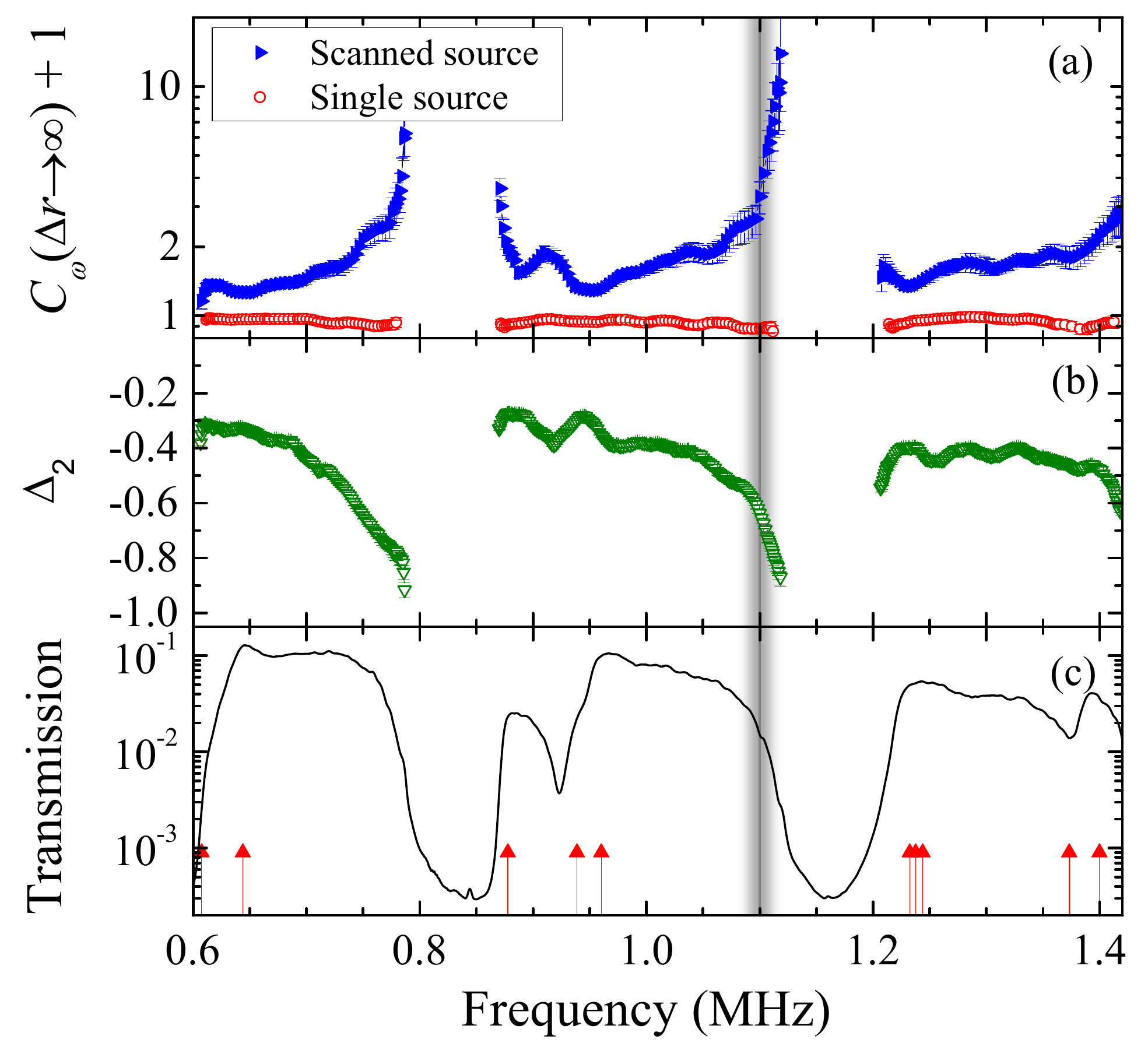}
\caption{\label{fig:c0andDelta2} (color online)
(a) Frequency dependence of the asymptotic value of the spatial correlation $C_{\omega}(\Delta r \to \infty) = C_0^{\text{(in)}}$ for single and scanned point sources, (b) anomalous dimension $\Delta_2$ of the gIPR for $q = 2$, and (c) the amplitude transmission coefficient. The arrows in (c) indicate the resonant frequencies of individual aluminum beads. $C_0^{\text{(in)}}$ and $\Delta_2$ show large increases in magnitude near the upper band edges, where we expect mobility edges to occur. The grey vertical line indicates the location of the ME near 1.1 MHz, estimated from transverse confinement measurements.}
\end{figure}

The frequency dependence of the asymptotic value of the spatial intensity correlation function between 0.6 and 1.4~MHz is shown in Fig.~\ref{fig:c0andDelta2}(a).
These data are the average of the measured correlations for $\Delta r$ between 25 and 50~mm,
where $C_{\omega}(\Delta r)$ is found to be independent of distance,
providing accurate measurements of $C_\omega(\Delta r \to \infty) = C_0^{\text{(in)}}$ when the source is scanned.
It increases rapidly with frequency near $0.78$ and $1.11$~MHz, reaching values up to 13 here, and even as high as 30 in other experiments---by far the largest values of $C_0$ ever reported. Comparison of these results with the amplitude transmission coefficient [Fig.~\ref{fig:c0andDelta2}(c)] reveals that the frequencies where $C_0$ increases rapidly coincide with the upper edges of pass bands in these disordered structures.
In the band gaps, the transmission becomes too small for long-range correlations to be measured. As explained in Refs.\ \cite{Turner1998a,Hu2008b}, these band gaps are not due to Bragg scattering, as in phononic crystals \cite{Yang2002}. Instead, they arise between pass bands formed from coupled resonances of the beads when the coupling is sufficiently weak.

Near the upper edges of the pass bands, where the average density of states decreases, mobility edges between extended and localized states may be expected \cite{John1987}.  Evidence that mobility edges do indeed occur near these band edges has been obtained through separate measurements of increased spatial confinement of the transmitted intensity near the upper band edges relative to the pass band centers, using the method developed by Hu \emph{et al.}\ \cite{Hu2008b}.  This evidence is most compelling for the ME near 1.1 MHz, which is indicated by the vertical line in Fig.~\ref{fig:c0andDelta2}.  Additional evidence can be inferred from the large increases that are found in the normalized intensity variance,
$C_{\omega}(\Delta r = 0, \Omega=0)$, near the upper band edges (e.g. Fig.~\ref{fig:sfcorr} and Ref. \cite{Hu2008b}).
Thus, the large increases in $C_0$ near $0.78$ and $1.1$ MHz must be due to large LDOS fluctuations near Anderson transitions in these samples, suggesting that $C_0$ is sensitive to critical effects.

This interpretation of the striking increase in $C_0$ near the band edges is further supported by measurements of the anomalous multifractal dimension $\Delta_2$, which characterizes the length-scale dependence of the inverse participation ratio (IPR) $P_2\sim L^{-d-\Delta_2}$ \cite{Faez2009a}.  The significant decrease in $\Delta_2$ near the upper band edges [Fig.~\ref{fig:c0andDelta2}(b)] is consistent with the expected behavior near the Anderson transition, where $\Delta_2$ should become increasingly negative, varying from $0$ in the diffuse regime to $-2$ deep in localized regime \cite{Evers2008a}.
Since the source and detector in our experiments are point-like, it is likely that a single mode dominates at any frequency, so we expect the IPR calculated from the intensity $I(\bf{r})$ and from the LDOS $\rho(\bf{r})$ to be equal \cite{Krachmalnicoff2010a,Murphy2011}.
Then, $P_2 = L^{-d}\langle \rho^2 \rangle / \langle \rho \rangle^2 = L^{-d}[C_0( \infty)+1]$, and we predict that
$\log\left[C_0\left(\infty \right) +1\right] \propto -\Delta_2$.
Within experimental error, the frequency dependencies of $C_{\omega}(\Delta r \to \infty) = C_0^{\text{(in)}}$ and $\Delta_2$ [Figs.~\ref{fig:c0andDelta2}(a) and (b)] are consistent with this prediction.
Thus, not only do the infinite-range correlations and the IPR show evidence of transitions from extended to localized behavior near the upper band edges, but the correspondence between these measurements verifies the link between $C_0$, $\Delta_2$, and LDOS fluctuations experimentally.

In conclusion, infinite-range intensity correlations have been measured directly in a strongly scattering 3D ``mesoglass'' for which Anderson localization of ultrasound was previously demonstrated \cite{Hu2008b}. Measurements are consistent with 
diagrammatic theory when large magnitudes of both long-range ($C_2$ and $C_3$) and infinite-range ($C_0$) terms are assumed. By varying the ultrasonic frequency, we have been able to investigate the growth not only of $C_2$ and $C_3$ but also of $C_0$ 
near the Anderson transition.  Infinite-range correlations of order unity are found over a broad range of frequencies, reflecting the high LDOS contrast that can be achieved in our samples. The magnitude of these $C_0$ correlations is seen to increase dramatically as a ME is approached and crossed. These $C_0$ results are mirrored by the frequency dependence of the anomalous dimension $\Delta_2$, which characterizes the size scaling of the inverse participation ratio. Our independent measurements of these two quantities establish a link between $C_0$ and $\Delta_2$, revealing that $C_0$ can be used to probe the Anderson transition. The possibility of exploiting our findings to experimentally investigate critical behavior at the Anderson transition, by focusing on the possible one-parameter scaling of $C_0$ near the ME, is a promising new avenue for future research.

This work was supported by NSERC and by a PICS program of CNRS.


\renewcommand{\vec}[1]{{\mathbf #1}}
\renewcommand{\thefigure}{S\arabic{figure}}
\renewcommand{\theequation}{S\arabic{equation}}
\renewcommand{\thetable}{S\arabic{table}}

\setcounter{equation}{0}
\setcounter{figure}{0}
\setcounter{table}{0}
\bibliographystyle{apsrev4-1}
\renewcommand*{\citenumfont}[1]{S#1}
\renewcommand*{\bibnumfmt}[1]{[S#1]}

\newpage
\begin{center}
\Large{\bf{Supplemental material}}
\end{center}
\normalsize







\section{Introduction}
This document provides a detailed description of how the experiments were performed and a summary of the theoretical results that were used to fit the data in the main text of the paper. Both the experimental procedure that was followed to conduct the ultrasonic measurements and the type of samples that were used are described.  For the theory, we compute the short-, long- and infinite-range correlation functions of intensity under assumption of weak disorder $k \ell \gg 1$, where $k$ is the wave number and $\ell$ is the mean free path due to disorder.

\maketitle
\section{Experimental details}
\label{secexpdet}

\subsection{Samples}
\begin{figure}
    \includegraphics*[width=8.5cm]{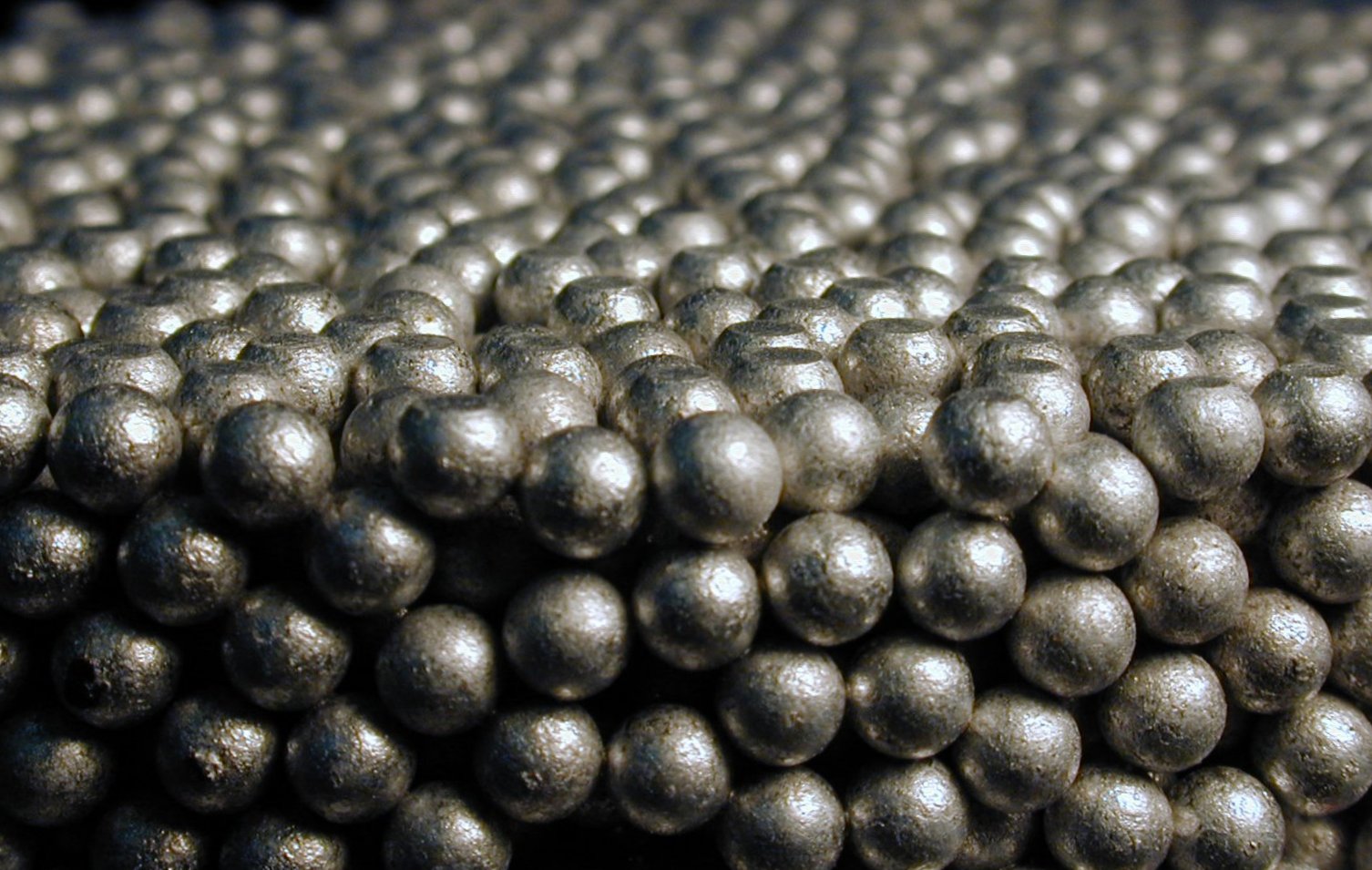}
	\caption{\label{fig:samppic} Photograph of one of the samples. Note the small ``necks'' connecting the beads, whose spherical shape is preserved.  The lightly polished top surface is also visible.}
\end{figure}

The samples investigated are disordered networks of aluminum beads, weakly brazed together to form disc-shaped slabs (see Fig.~\ref{fig:samppic}).
The beads used are monodisperse and $4.11\pm 0.03$~mm in diameter, and the samples have a volume fraction of approximately 55\%, consistent with random loose packing. The beads were weakly bonded together by precisely controlling the flux, alloy concentration, and temperature during brazing, such that the spherical bead structure of the individual beads remained intact, with only small necks elastically connecting the beads.
The samples were thoroughly cleaned to remove any surface contaminants from the beads that could lead to spurious dissipation in the ultrasonic experiments.  The front and back surfaces of the samples were lightly polished to ensure that the opposite faces of the slabs were flat and parallel. The slabs were 120~mm in diameter, much larger than the sample thicknesses, in order that the slabs be sufficiently wide to avoid edge effects.  Anderson localization has been observed in these samples for thicknesses $L$ ranging from 8.3 to 23.5~mm \cite{natphysloc}.  While correlation measurements were also performed for several sample thicknesses, the results shown in this paper are all for $L = 14.5$~mm, which was a representative data set for which the most complete results were obtained.

These samples exhibit very strongly scattering of ultrasound in the frequency range of the experiments, as we have determined by measuring the weak coherent signal that propagates ballistically through the sample \cite{page1996,natphysloc}.  Representative results of these measurements are shown in Table~\ref{tab:params}.  At all frequencies, the scattering mean free path $\ell_s$ is considerably smaller than both the diameter of a single bead and the measured wavelength $\lambda$ inside the samples, with the product of wave vector and mean free path $k\ell_s$ being of order unity.
\begin{table}
\caption{\label{tab:params}
Experimentally determined parameters.  The phase velocity, scattering mean free path, wavelength, and scattering strength are determined from ballistic measurements.  The focal spot size was measured by scanning the hydrophone detector in the source plane.
}
\centering
\begin{ruledtabular}
\begin{tabular}{ccccc}
 Frequency &~0.6 MHz~&~1.0 MHz~&~1.4 MHz~&~2.4 MHz~\\ [0.3ex]
\colrule
\multicolumn{5}{c}{Ballistically measured parameters}\\ [0.3ex]
\colrule
$\ell_s$ (mm)& 1.0 & 0.7 & 0.8 & 0.6 \\[0.3ex]
$v_p$ (mm/$\upmu$s)& 2.7 & 2.8 & 2.8 & 5.0 \\[0.3ex]
$\lambda$ (mm) & 4.5 & 2.8 & 2.0 & 2.1 \\[0.3ex]
$k\ell_s$ & 1.4 & 1.7 & 2.5 & 1.8 \\[0.3ex]
\colrule
\multicolumn{5}{c}{Focal spot size}\\ [0.3ex]
\colrule
FWHM (mm) & 1.5 & 1.2 & 0.93 & 0.91 \\[0.3ex]
\end{tabular}
\end{ruledtabular}

\end{table}
\subsection{Measurement procedures}

\begin{figure}
    \includegraphics*[width=8.5cm]{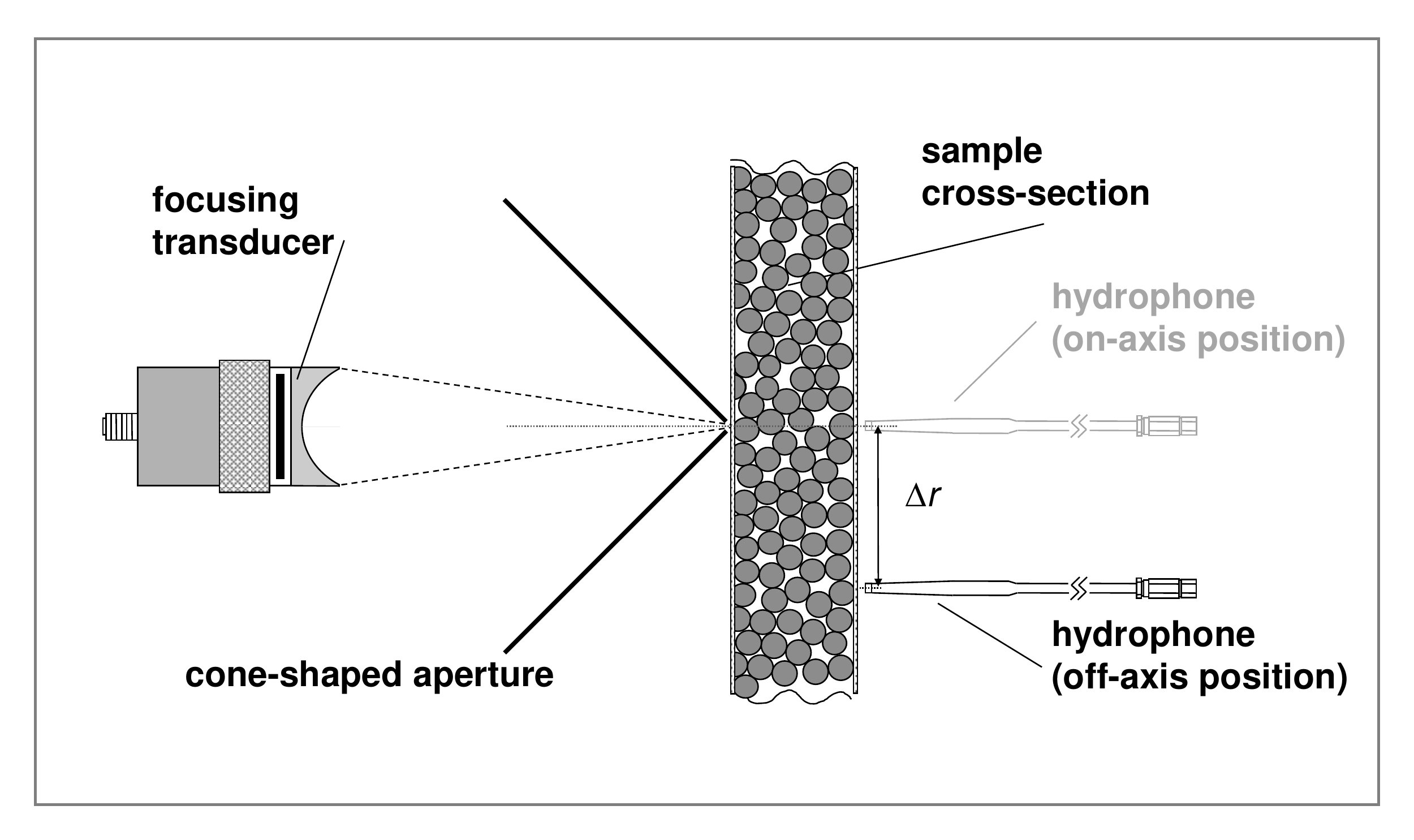}
	\caption{\label{fig:expsetup} Schematic diagram showing experimental setup.}
\end{figure}

The experiments were performed in a water tank to capitalize on the flexibility of ultrasonic immersion transducer technology for controlling source and detector positions.   A focused ultrasonic pulse was incident on the sample, and the transmitted waves were measured in the near field on the opposite side using a miniature hydrophone detector.  A schematic representation of the experimental configuration is shown in Fig.~\ref{fig:expsetup}.
	
Because we are interested in measuring ultrasonic transport through the solid elastic network of aluminum beads, 
the samples were mounted into acrylic holders and sealed with thin plastic walls to prevent water from entering the pore space surrounding the beads; hence, within the pores, only air (for the measurements at 2.4 MHz) or vacuum (for the lower frequency measurements between 0.5 and 1.5 MHz) was present.  To ensure good acoustic coupling between the front and back sample surfaces and the flat waterproofing walls, the walls were coated with a very thin layer of an ultrasonic couplant.

The ultrasonic pulse was generated by focusing immersion transducers, which have front surfaces that are curved to act as a lens.  The central frequencies of the transducers were either 1.0 or 2.25~MHz. The transducers were designed to have a focal length of approximately 30~cm, and a conical screen with a small aperture was placed at the focus to remove any side lobes or other beam artifacts. 
This large focal distance was selected to enable the multiply scattered, transmitted signals to be recorded before the arrival of any spurious echoes that had reverberated back and forth between the transducer and sample.  Note also that this large focal distance ensured that the source was temporally decoupled from the sample, since the time interval between the emission of the pulse at the transducer and its arrival at the sample surface is very much longer than the incident pulse width.  Thus, the pulse incident on the sample surface was a constant amplitude pressure pulse, with magnitude and bandwidth that was independent of the LDOS at the focal spot.  This type of source has the advantage of simplicity for investigating the effect of LDOS fluctuations at the input surface on the intensity correlations of the transmitted signals,  although its nature is quite different to sources in optics that have been used to investigate LDOS fluctuations themselves via the strong coupling that exists in the photonic environment around the sources and scatterers \cite{classsource}.  In our experiments, for each source/detector location, the pulse was repeated several thousand times at a repetition rate of several hundred Hertz (slow enough to ensure that all signals due to the previous pulse had died completely away), so that the recorded signals could be averaged to improve the signal-to-noise ratio.

The conical screen was wrapped in Teflon tape to make it acoustically opaque.
The cone shape was chosen so that edges of the focused beam could be effectively blocked when the aperture was placed close to the sample, while at the same time preventing significant stray sound being reflected back towards the sample from the screen.  The pressure field at the source plane (about 1~mm from the aperture) was mapped using the hydrophone detector, so that the spatial extent of the source spot on the sample surface could be determined.  The recorded signals were Fourier transformed and the intensity maps at each frequency were fit to a Gaussian in order to determine the source size.  The results of these measurements are shown in Table~\ref{tab:params}.  Note that the focal spot size at each frequency is significantly less than the wavelength inside the sample, and comparable to one wavelength in water.

The hydrophone used in these experiments is a sub-wavelength phase-sensitive detector with an active element diameter of $400~\upmu$m.  The hydrophone has a needle-like shape, which serves to minimize reflections back to the sample.  In our experiments, the hydrophone was placed approximately 1~mm from the sample surface (less than one wavelength in water), and thereby records the near-field transmission with good spatial resolution.

The intensity correlations of the transmitted ultrasonic waves were determined by first taking the Fourier transform of the recorded pressure fields $p(\vec{r},t)$ and squaring the magnitude of the Fourier transforms to obtain signals that are proportional to the ultrasonic intensity at each frequency. Two types of experiments were performed in order to isolate the contributions to the intensity correlations of fluctuations in the local density of states at the source positions. The first set of experiments was designed to directly measure the $C_0$ correlations due to these LDOS fluctuations at the source.  In these experiments, the transmitted signals were recorded at 13 detector positions for each source location.  In order to get good statistics, the sample was scanned to have the source focused on over 3000 independent locations. For each pair of detector positions, the correlations were calculated for all source locations, and the results of the correlations for similar values of $\Delta r$ were binned and averaged together.
In the second set of experiments, which were designed to suppress these $C_0$ correlations, over 3000 detector positions were used for a single source location.  Correlations were calculated for every possible pair of detector positions, and results for similar values of $\Delta r$ were again binned.  This experiment was repeated for seven independent source locations, and the results of each of these experiments were averaged together.

\section{Theory for spatial correlations}
\label{secspatial}

\subsection{Definitions}

We consider the spatial correlation function of intensity fluctuations $\delta I(\vec{r}, \omega) = I(\vec{r}, \omega) - \langle I(\vec{r}, \omega) \rangle$:
\begin{eqnarray}
C_{\omega}(\vec{r}, \vec{r}') = \frac{\langle \delta I(\vec{r}, \omega)
\delta I(\vec{r}', \omega) \rangle}{\langle I(\vec{r}, \omega) \rangle
\langle I(\vec{r}', \omega) \rangle}.
\label{cdef}
\end{eqnarray}
To lighten the notation, we will omit the subscript `$\omega$' from here on, keeping in mind that all measurements are performed for waves at the same frequency $\omega$. Typically, $C(\vec{r}, \vec{r}')$ is a decaying function of $\Delta r = |\vec{r}-\vec{r}'|$ with $C(\vec{r}, \vec{r}' = \vec{r}) = \langle \delta I(\vec{r})^2 \rangle/\langle I(\vec{r}) \rangle^2$ being the normalized variance of intensity, and $C(\vec{r}, \vec{r}') = 0$ in the absence of intensity correlations. It is convenient to split $C(\vec{r}, \vec{r}')$ in several parts: $C = C_1 + C_2 + C_3 + C_0$, each of $C_i$ originating from different physical processes \cite{akker07, shapiro99}.

\subsection{Short-range correlation $C_1$}

In the bulk of a disordered medium and far from boundaries, the short-range contribution to $C$ is \cite{shapiro86,akker07}
\begin{eqnarray}
C_1^{\mathrm{bulk}}(\vec{r}, \vec{r}') =
\left( \frac{\sin k \Delta r}{k \Delta r} \right)^2 \exp(-\Delta r/\ell),
\label{c1bulk}
\end{eqnarray}
where $k = 2\pi/\lambda$, $\ell$ is the scattering mean free path.
At the surface of a disordered sample, the spatial correlation $C_1$ is modified due to the anisotropic angular distribution of intensity \cite{freund92}:
\begin{eqnarray}
C_1^{\mathrm{surface}}(\vec{r}, \vec{r}') &=&
\left\{\frac{1}{\Delta + \frac12} \left[\Delta \frac{\sin k \Delta r}{k \Delta r} + \frac{J_1(k \Delta r)}{k \Delta r}\right]
\right\}^2
\nonumber \\
&\times& \exp(-\Delta r/\ell) = h(k \Delta r, k \ell),
\label{c1surface}
\end{eqnarray}
where $\Delta = z_0/\ell^*$ with $z_0$ the extrapolation length entering the boundary conditions for the average intensity,
and we defined the function $h(k\Delta r, k\ell)$ that will be used in the following. This result is largely independent of the spatial extent of the source (plane wave, beam of finite size or point source) and has been tested experimentally \cite{sebbah00}.

The $C_1$ intensity correlation is equal to the square of the field correlation function, which we can measure directly since our detector records the transmitted pressure field.
Our least-squares fits of the square root of Eq.~(\ref{c1surface}) to our experimental field correlation data yield values of the parameters $k$ and $\ell$ that are consistent with those obtained from ballistic measurements \cite{page1996}.  This indicates that Eq.~(\ref{c1surface}), with $k$ and $\ell$ taken from ballistic measurements, gives a good description of our experimental results, which were obtained from measurements performed just outside the sample, within one wavelength of its surface.
Thus, Eq.~(\ref{c1surface}) gives a reliable characterization of our experimental data for the $C_1$ contribution to the intensity correlation, supporting our use of this expression in our analysis of the total correlation $ C_{\omega}(\vec{r}, \vec{r}')$.

A more comprehensive study of the $C_1$ correlation would involve taking into account near-field effects in a way similar to a recent analysis published for electromagnetic waves \cite{carminati2010}. In any case, the precise form of $C_1$ does not play an important role in our analysis, which is mainly focused on the long-range correlations that are described next. We thus postpone a detailed analysis of $C_1$ to a future publication.


\subsection{Long-range correlation $C_2$}

In contrast to $C_1$, the long-range contribution $C_2$ depends on the spatial extent of the source. It is not easy to calculate for an arbitrary source. In addition, we have to make an assumption of weak disorder ($k \ell \gg 1$) to compute the diagrams corresponding to $C_2$.

{\bf Transmission of a plane wave through a slab.}
We assume that a slab of thickness $L \gg \ell$ and transverse extent $W \gg L$ is illuminated by a plane wave. The spatial correlation of intensity is calculated at the opposite side of the slab, as a function of transverse distance $\Delta r = |\mathbf{r} - \mathbf{r}'| \gg \ell^*$ \cite{stephen87, pnini89}:
\begin{eqnarray}
C_2^{\mathrm{plane\; wave}}(\Delta r) &=& \frac{3}{2 (k \ell^*)^2} \frac{\ell^*}{L}
\left[\frac{L}{\Delta r} + F\left(\frac{\Delta r}{L} \right) \right]
\nonumber \\
&\simeq&
\begin{cases}
\frac{3}{2 (k \ell^*)^2} \frac{\ell^*}{\Delta r}, &\Delta r \ll L,\cr
\propto e^{-\pi \Delta r/L}, &\Delta r > L,
\end{cases}
\label{c2plane}
\end{eqnarray}
where
\begin{eqnarray}
F(x) = \frac12 \int_0^{\infty} dq J_0(qx)
\left( \frac{\sinh 2q - 2q}{\sinh^2 q} - 2 \right),
\label{f}
\end{eqnarray}
and $\ell^*$ is the transport mean free path.

\begin{figure}[h]
\hspace*{-1.0cm}
\includegraphics[width=1.2\columnwidth]{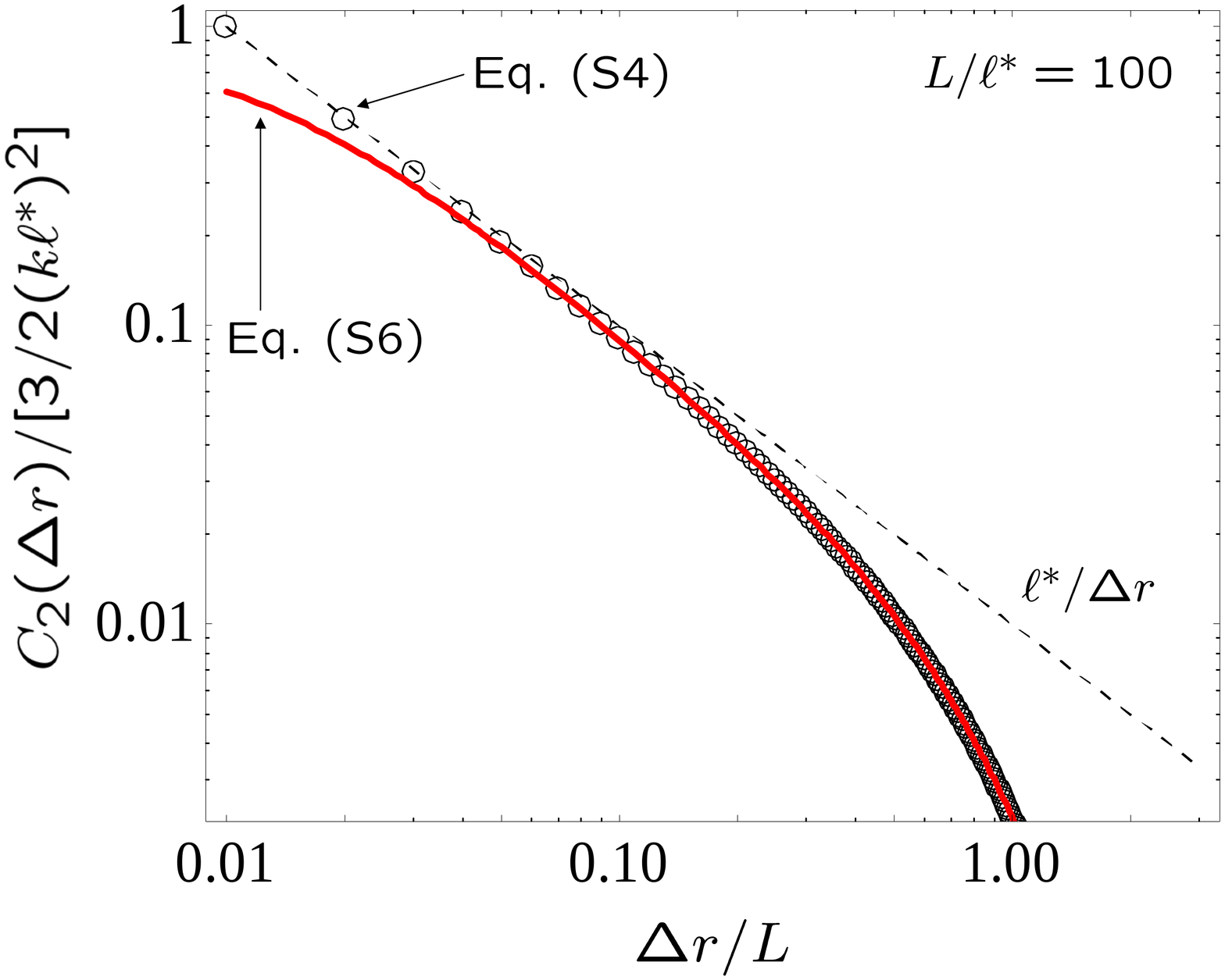}
\vspace*{-1cm}
\caption{Comparison of Eqs.\ (\ref{c2plane}) (shown by circles) and (\ref{c2plane2}) (shown by the solid red line) for the long-range correlation function of intensity fluctuations. The dashed line shows $\ell^*/\Delta r$.}
\label{figplane}
\end{figure}

Equation (\ref{c2plane}) applies for $\Delta r \gg \ell^*$, where it exhibits the interesting slow decay, which is why this correlation function is often referred to as ``long-range''. However, the physical processes giving rise to this behavior are at work for $\Delta r \lesssim \ell^*$ as well.
For $\Delta r = 0$, for example, they contribute to the variance of the intensity fluctuations $\langle \delta I(\vec{r})^2 \rangle/\langle I(\vec{r}) \rangle^2$.
Physically, we expect $C_2(\Delta r = 0) \simeq C_2(\Delta r = \ell^*)$, but Eq.\ (\ref{c2plane}) diverges for $\Delta r \to 0$. This divergence is an artifact of approximations made during the derivation of Eq.\ (\ref{c2plane}).
A more precise shape of $C_2(\Delta r)$ at small $\Delta r \lesssim \ell^*$ can be obtained by paying more attention to large $q$ and avoiding the limit $q \ell^* \ll 1$ which is tacitly taken in the derivation of Eq.\ (\ref{c2plane}). We then obtain a longer but more accurate expression for $C_2$:
\begin{eqnarray}
C_2^{\mathrm{plane\; wave}}(\Delta r) = \frac{3}{2 (k \ell^*)^2} F_2\left(\frac{\Delta r}{L},\frac{\ell^*}{L} \right),
\label{c2plane2}
\end{eqnarray}
where \begin{eqnarray}
F_2(x,y) &=& \frac{y}{2} \int_0^{\infty} du
\frac{J_0(ux)}{(u y \sinh u)^2}
\left\{
\sinh^2(u y)
\right.
\nonumber \\
&\times& \left.
\left[\sinh[2 u (1-y)] - 2 u (1-y) \right]
\right.
\nonumber \\
&+& \left. \sinh^2[u (1-y)] \left[\sinh(2 u y) - 2 u y \right]
\right\}
\nonumber \\
&\simeq&
\begin{cases}
1-y, &x = 0, \cr
y/x, &y \ll x \ll 1,\cr
\propto e^{-\pi x}, &x > 1.
\end{cases}
\label{ftilde}
\end{eqnarray}
A comparison of Eqs.\ (\ref{c2plane}) and (\ref{c2plane2}) is shown in Fig.\ \ref{figplane}. The latter equation, in contrast to Eq.\ (\ref{c2plane}), allows us to obtain the value of $C_2$ for $\Delta r = 0$:
\begin{eqnarray}
C_2^{\mathrm{plane\; wave}}(\Delta r = 0) = \frac{3}{2 (k \ell^*)^2} \left( 1 - \frac{\ell^*}{L} \right).
\label{c2zero}
\end{eqnarray}

{\bf Point source in the infinite medium.}
Even if this geometry seems simple, the calculation of $C_2$ appears quite involved. If we define the center of mass $\mathbf{R} = \frac12 (\vec{r}_1 + \vec{r}_2)$ and the difference
$\Delta \mathbf{r} = \vec{r}_1 - \vec{r}_2$ coordinates, we can obtain simple results for $\Delta \vec{r} \perp \vec{R}$:
\begin{eqnarray}
C_2^{\mathrm{point\; source}}(\Delta r) \simeq \pm
\frac{3}{2 (k \ell^*)^2} \frac{\ell^*}{\Delta r},
\label{c2point}
\end{eqnarray}
with the `$+$' sign for $\Delta r \ll R$ and the `$-$' sign for $\Delta r \gg R$.

{\bf Transmission of a tightly focused beam through a slab.}
This situation is realized in our experiments and is somewhat intermediate with respect to the two previous cases (the sample is a slab, but the source is point-like). Because the results for the plane wave incident on a slab (\ref{c2plane2}) and the point source in the infinite medium (\ref{c2point}) coincide  for $\ell^* < \Delta r < L$, we expect that the same result will also hold for the tightly focused beam. We will therefore use:
\begin{eqnarray}
C_2^{\mathrm{focused\; beam}}(\Delta r) &\simeq&
\frac{3}{2 (k \ell^*)^2} F_2\left(\frac{\Delta r}{L},\frac{\ell^*}{L} \right),
\nonumber \\
&&\Delta r < L.
\label{c2focus}
\end{eqnarray}
For $\Delta r > L$, we expect $C_2^{\mathrm{focused\; beam}}(\Delta r)$ to be different from both Eqs.\ (\ref{c2plane2}) and (\ref{c2point}), but because its magnitude is already small at such large distances, it will not play a significant role in the fits to the experimental data.

{\bf Short-range part of $C_2$}.
The calculation leading to Eqs.\ (\ref{c2plane2}), (\ref{c2point}) and (\ref{c2focus}) also yields short-range terms that are rarely mentioned but exist. The full expression for $C_2$ including both long- and short-range contributions is
\begin{eqnarray}
C_2^{\mathrm{full}}(\Delta r) &\simeq&
\frac{3}{2 (k \ell^*)^2}
\left[ F_2\left(0,\frac{\ell^*}{L} \right) h(k\Delta r, k\ell) \right.
\nonumber \\
&+& \left. F_2\left(\frac{\Delta r}{L},\frac{\ell^*}{L} \right) \right].
\label{c2full}
\end{eqnarray}

\subsection{Long-range correlation $C_3$}

The calculation of the spatial $C_3$ correlation function for a beam focused on the surface of a 3D disordered slab is a complicated task that we did not succeed in accomplishing. However, the structure of the result may be anticipated from the diagrams involved in the calculation \cite{akker07}: we expect short- and long-range terms similar to $C_2$. The magnitude of $C_3$ is expected to be of order $1/(k \ell^*)^4$. Hence, an approximate expression for $C_3$ may be written as
\begin{eqnarray}
C_3(\Delta r) &\simeq&
\frac{\mathrm{const}}{(k \ell^*)^4}
\left[ F_2\left(0,\frac{\ell^*}{L} \right) h(k\Delta r, k\ell) \right.
\nonumber \\
&+& \left. F_2\left(\frac{\Delta r}{L},\frac{\ell^*}{L} \right) \right].
\label{c3}
\end{eqnarray}

\subsection{Infinite-range correlation $C_0$}

\begin{figure}[t]
\hspace*{-1.6cm}
\includegraphics[width=1.5\columnwidth]{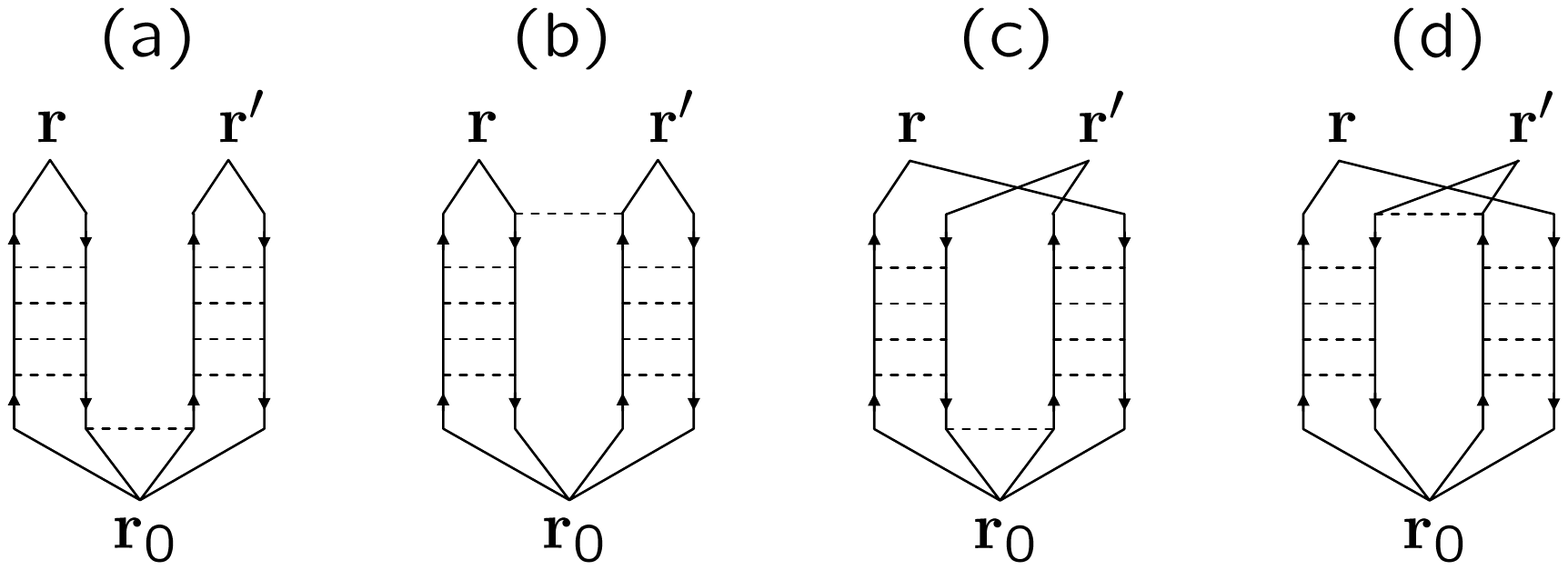}
\vspace*{-6cm}
\caption{Diagrams contributing to $C_0$ correlation function. $\vec{r}_0$ is the source position. The diagram (a) is the original long-range one \cite{shapiro99}; it is independent of $\Delta r = |\vec{r}-\vec{r}'|$. The diagram (b) is short-range. It was calculated in Ref.\ \cite{retzker02}. The diagrams (c) and (d) are both short-range and were not considered previously. A complex conjugate diagram should be added to each of the diagrams.}
\label{diagramsc0}
\end{figure}

Similarly to $C_2$ and $C_3$, $C_0$ correlation also contains the `interesting', infinite-range part and the `trivial', short-range one. The full expression is found by summing the diagrams of Fig.\ \ref{diagramsc0}:
\begin{eqnarray}
C_0^{(a)}(\Delta r) &=&
C_0^{\mathrm{(in)}},
\label{c0a}
\\
C_0^{(b)}(\Delta r) &=&
C_0^{\mathrm{(out)}}
\frac{f_b(k \Delta r, k\ell)}{f_b(0, k\ell)},
\label{c0b}
\\
C_0^{(c)}(\Delta r) &=&
C_0^{\mathrm{(in)}} h(k \Delta r, k\ell),
\label{c0c}
\\
C_0^{(d)}(\Delta r) &=&
C_0^{\mathrm{(out)}}
\frac{f_d(k \Delta r, k\ell)}{f_d(0, k\ell)},
\label{c0d}
\end{eqnarray}
where $C_0^{\mathrm{(in)}}$ is the genuine, infinite-range correlation that survives at large $\Delta r$ \cite{shapiro99}. It results from the scattering near the source and is related to the variance of the local density of states at $\vec{r}_0$ \cite{vantiggelen06}. In contrast, the terms (\ref{c0b}) and (\ref{c0d}), which are proportional to $C_0^{\mathrm{(out)}}$, result from the scattering near the detection points $\vec{r}_1$ and $\vec{r}_2$. They are appreciable only at small $\Delta r = |\vec{r}_1 - \vec{r}_2|$. For the white-noise uncorrelated disorder, $C_0^{\mathrm{(in)}} = C_0^{\mathrm{(out)}} = \pi/k\ell$. In our experiment, the disorder is correlated and the symmetry between the `point-like' excitation and the `point-like' detection may be broken because neither is actually point-like and the effective sizes of the excitation and detection areas may differ, so that $C_0^{\mathrm{(in)}} \ne C_0^{\mathrm{(out)}} \ne \pi/k\ell$. Moreover, these parameters are not universal and will depend on the microscopic structure of the disordered sample \cite{skip00}. We use $C_0^{\mathrm{(in)}}$ and $C_0^{\mathrm{(out)}}$ as free fit parameters when comparing theory to the experimental data.

\begin{figure}[t]
\hspace*{-1cm}
\includegraphics[width=2\columnwidth]{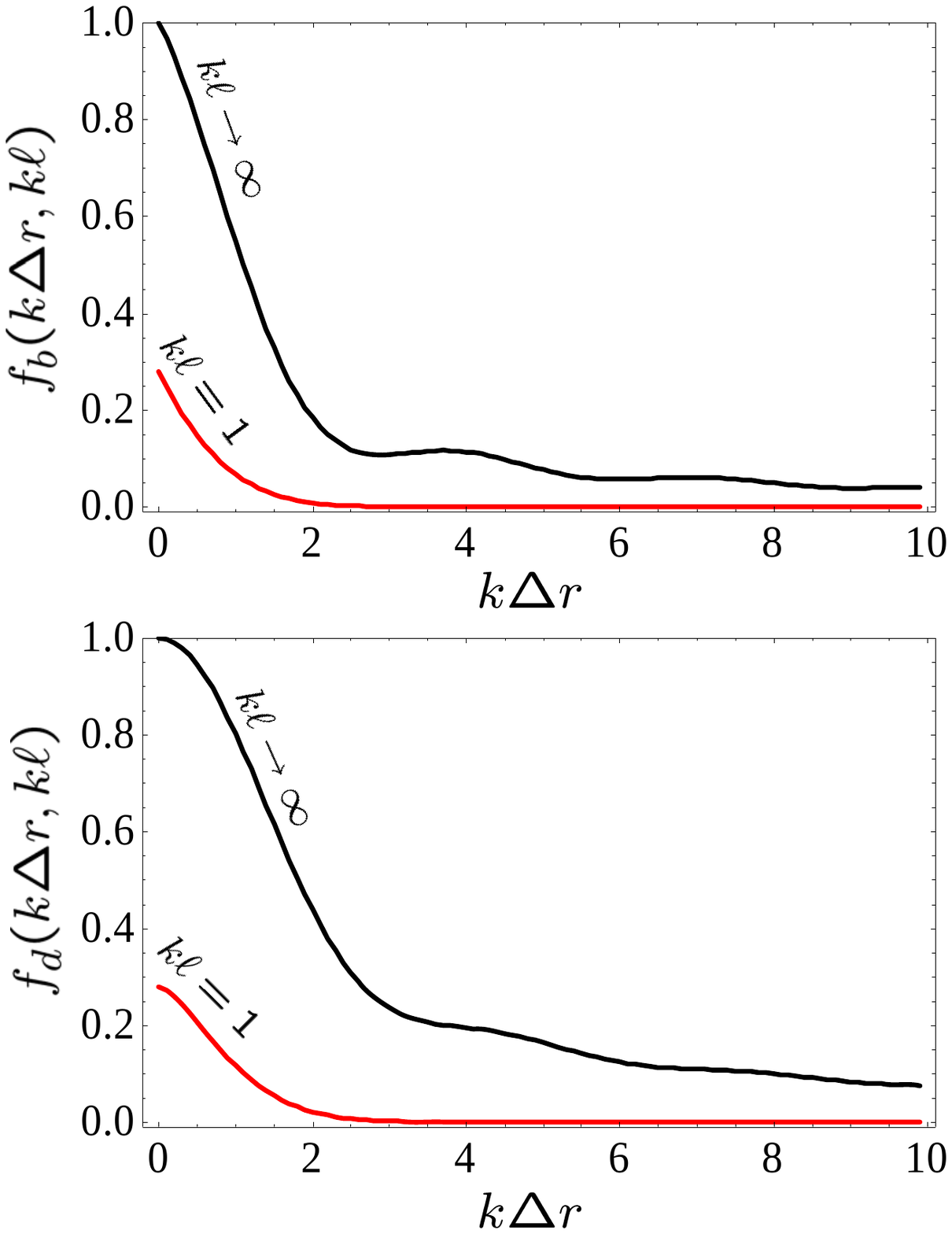}
\vspace*{-1.5cm}
\caption{Functions $f_b$ and $f_d$ describing the short-range part of $C_0$ correlation function.}
\label{f2f4}
\end{figure}

The functions $f_b(k \Delta r, k\ell)$ and $f_d(k \Delta r, k\ell)$ are rapidly decaying functions of $k\Delta r$:
\begin{eqnarray}
f_b(k\Delta r, k\ell) &=& \frac{1}{2\pi k \Delta r}
\mathrm{Re}
\left\{
i \int_0^{\infty} dx \frac{\sin x}{x} e^{-(i + 1/k\ell) x}
\right.
\nonumber \\
&\times& \left.
\Big(
\mathrm{Ei}[-(k\Delta r + x)/k\ell]
\right.
\nonumber \\
&-& \left.
\mathrm{Ei}[(2i - 1/k\ell)(k\Delta r + x)]
\right.
\nonumber \\
&+&
\left.
\mathrm{Ei}[(2i - 1/k\ell)|k\Delta r - x|]
\right.
\nonumber \\
&-& \left.
\mathrm{Ei}[-|k\Delta r - x|/k\ell]
\Big)
\vphantom{\int_0^{\infty}}\right\},
\label{fb}
\\
f_d(k\Delta r, k\ell) &=& \frac{1}{\pi k \Delta r}
\int_0^{\infty} dx \frac{\sin^2 x}{x} e^{-x/k\ell}
\nonumber \\
&\times&
\left[
\mathrm{Ei}\left( -\frac{k\Delta r + x}{k\ell} \right)
\right.
\nonumber \\
&-& \left.
\mathrm{Ei}\left( -\frac{|k\Delta r - x|}{k\ell} \right)
\right].
\label{fd}
\end{eqnarray}
The behavior of these functions is illustrated in Fig.\ \ref{f2f4}.

\subsection{Full expression for the correlation function}

Adding up Eqs.\ (\ref{c1surface}), (\ref{c2full}), (\ref{c3}) and (\ref{c0a})--(\ref{c0d}) we end up with an expression that can be used to fit experimental data:
\begin{eqnarray}
C(\Delta r) &=&
\left[ 1 + A + C_0^{\mathrm{(in)}} \right] h(k\Delta r, k\ell)
\nonumber \\
&+& A \times \frac{F_2\left(\Delta r/L, \ell^*/L \right)}{F_2\left(0, \ell^*/L \right)}
+ C_0^{\mathrm{(in)}}
\nonumber \\
&+& C_0^{\mathrm{(out)}} \left[ \frac{f_b(k \Delta r, k\ell)}{f_b(0, k\ell)} + \frac{f_d(k \Delta r, k\ell)}{f_d(0, k\ell)}
\right],\;\;\;
\label{cfull}
\end{eqnarray}
where $A = [3/2 (k\ell^*)^2 + \mathrm{const}/(k \ell^*)^4] F_2\left(0, \ell^*/L \right)$. $A$, $C_0^{\mathrm{(in)}}$ and $C_0^{\mathrm{(out)}}$ are the unknown fit parameters.

\subsection{Role of the finite beam waist}
\label{source}
In our experiments, the beam of ultrasound is focused to a small spot of size $w < \lambda$ on the sample surface, whereas theoretical results to which we compare our measurements are obtained for either an incident plane wave ($C_1$, $C_2$ and $C_3$) or a point source ($C_1$, $C_2$, $C_3$ and $C_0$). For the long-range correlations $C_2$ and $C_3$, this leads to the need of adjusting the parameter $A$ to account empirically for the finite beam width
when the comparison to experiments is made. Since $A$ decreases as $w$ increases
\cite{pnini89}, our measurements underestimate the magnitude of the $C_2$ and $C_3$ correlations that would be measured for a true point-like source of waves.
For the infinite-range $C_0$ correlation, having a source of finite size is known to reduce the magnitude of $C_0^{\mathrm{(in)}}$ \cite{skip00}, which is not exactly equal to the variance of LDOS anymore and will depend on $w$ as well. Therefore, our measurements of $C_0$ underestimate the actual LDOS fluctuations, which
are expected to be even stronger. However, the analysis presented in the main text of the paper does not rely on the magnitude of the correlations measured experimentally, but investigates
their dependence on the distance between measurement points [Figs.\ \ref{fig:spatcorr1} and \ref{fig:sfcorr}(a)], the frequency difference [Fig.\ \ref{fig:sfcorr}(b)], or the central frequency (Fig.\ \ref{fig:c0andDelta2}). As a consequence, our conclusions remain valid independent of the beam waist $w$. A more detailed, quantitative comparison between theory and experiment would require calculating the dependence of $C_2$, $C_3$ and $C_0$ on $w$, so that the number of free parameters would be reduced
in the fits presented in Figs.\ \ref{fig:spatcorr1} and \ref{fig:sfcorr}. However, such an analysis is beyond the scope of the present work and is not required to arrive at the conclusions that we make in the main text of the Letter.

\subsection{Accounting for the size of the detector}
\label{size}

In the experiment, the acoustic field is measured very close to the sample surface with a disk-shaped hydrophone of radius $b = 0.2$ mm. In order to take into account the size of the hydrophone, we assume that the measured quantity is not the intensity $I(\vec{r})$ at a point $\vec{r}$ but the intensity averaged over a disk of radius $b$ centered at $\vec{r}$:
\begin{eqnarray}
{\cal I}(\vec{r}) = \frac{1}{\pi b^2} \int_{b(\vec{r})} I(\vec{r}') d^2 \vec{r}',
\label{intint}
\end{eqnarray}
where $b(\vec{r})$ denotes a disk of radius $b$ centered at $\vec{r}$. The correlation function of ${\cal I}(\vec{r})$ can be then obtained from the correlation of $I(\vec{r})$ by a double spatial integration:
\begin{eqnarray}
C_{\cal I}(\Delta\vec{r} &=& \vec{r}_1 - \vec{r}_2) = \frac{1}{(\pi b^2)^2} \int_{b(\vec{r}_1)} d^2 \vec{r}_1' \int_{b(\vec{r}_2)} d^2 \vec{r}_2'
\nonumber \\
&\times& C_I(\Delta\vec{r}' = \vec{r}_1' - \vec{r}_2').
\label{corrcorr}
\end{eqnarray}

\section{Theory for frequency correlations}

\subsection{Definitions}

The frequency correlation function of intensity fluctuations $\delta I(\vec{r}, \omega) = I(\vec{r},\omega) - \langle I(\vec{r}, \omega) \rangle$ is defined as
\begin{eqnarray}
C_{\omega}(\vec{r}, \Omega) = \frac{\langle \delta I(\vec{r}, \omega + \frac12 \Omega)\;
\delta I(\vec{r}, \omega-\frac12 \Omega) \rangle}{\langle I(\vec{r}, \omega) \rangle^2},
\label{cdefomega}
\end{eqnarray}
where we assume that the average intensity is independent of frequency in the frequency band under consideration: $\langle I(\vec{r}, \omega + \frac12 \Omega) \rangle = \langle I(\vec{r}, \omega - \frac12 \Omega) \rangle = \langle I(\vec{r}, \omega) \rangle$. Once again, we will omit the subscript `$\omega$' of $C$ from here on. The behavior of $C(\vec{r}, \Omega)$ with $\Omega$ is similar to the behavior of the spatial correlation function $C(\Delta r)$ with $\Delta r$: it decays and has both short- and long-range parts.

\subsection{Short-range correlation $C_1$}

The short-range part of $C$ can be easily calculated in transmission of a plane wave through a slab of thickness $L$ \cite{akker07,sebbah00}:
\begin{eqnarray}
C_1(\Omega) = \left|\frac{L}{\ell^*} \times \frac{\sinh^2(\alpha \ell^*)}{\alpha \ell^* \sinh \alpha L} \right|^2,
\label{c1slabomega}
\end{eqnarray}
where
$\alpha L = \pi \sqrt{i \Omega/\Omega_{\mathrm{Th}}}$,
$\Omega_{\mathrm{Th}} = \pi^2 D/L^2$ is the Thouless frequency, $D$ is the diffusion coefficient of the wave, and we neglected corrections due to boundary conditions, assuming $L \gg \ell^*$, $z_0$.

For a point source at the origin in the infinite medium we have
\begin{eqnarray}
C_1(\vec{R}, \Omega) = \left|
\exp\left( -\alpha R \right) \right|^2,
\label{c1infomega}
\end{eqnarray}
with similar definitions
$\alpha R = \pi \sqrt{i \Omega/\Omega_{\mathrm{Th}}}$, $\Omega_{\mathrm{Th}} = \pi^2 D/R^2$.

Both correlation functions (\ref{c1slabomega}) and (\ref{c1infomega}) oscillate and decay roughly exponentially with $\sqrt{\Omega/\Omega_{\mathrm{Th}}}$, so that no correlation is left for $\Omega \gg \Omega_{\mathrm{Th}}$. We will adopt Eq.\ (\ref{c1slabomega}) in the following.

\subsection{Long-range correlation $C_2$}

{\bf Transmission of a plane wave through a slab.}
From a calculation following Refs.\ \cite{pnini89,deboer92} we found the following result:
\begin{eqnarray}
C_2^{\mathrm{plane\; wave}}(\Omega) =
\frac{3}{2(k \ell^*)^2}
F_2 \left( \alpha L, \frac{\ell^*}{L} \right),
\label{c2slabomega}
\end{eqnarray}
where
\begin{eqnarray}
F_2 \left( \alpha L, \frac{\ell^*}{L} \right) &=&
\frac12 \frac{L}{\ell^*}
\int\limits_0^{\infty} du\; u\; f \left( \frac{u}{L}, \alpha L, \frac{\ell^*}{L} \right),
\label{ftildeomega}
\\
f\left(q, \alpha L, \frac{\ell^*}{L} \right) &=&
\frac{4}{L}
\left[
\int\limits_0^{\ell^*} dz
\left(
\frac{\sinh qz\; \sinh q\ell^*}{q \sinh qL}
\right. \right.
\nonumber \\
&\times& \left. \left.
\left|
\frac{\partial}{\partial z}
\frac{\sinh \alpha z\; \sinh \alpha (L-\ell^*)}{\alpha \ell^*\; \sinh \alpha L}
\right|
\right)^2
\right.
\nonumber \\
&+& \left.
\int\limits_{\ell^*}^{L-\ell^*} dz
\left(
\frac{\sinh qz\; \sinh q\ell^*}{q \sinh qL}
\right. \right.
\nonumber \\
&\times& \left. \left.
\left|
\frac{\partial}{\partial z}
\frac{\sinh \alpha \ell^*\; \sinh \alpha (L-\ell^*)}{\alpha \ell^*\; \sinh \alpha L}
\right|
\right)^2
\right.
\nonumber \\
&+& \left.
\int\limits_{L-\ell^*}^L dz
\left(
\frac{\sinh q(L-\ell^*)\; \sinh q(L-z)}{q \sinh qL}
\right. \right.
\nonumber \\
&\times& \left. \left.
\left|
\frac{\partial}{\partial z}
\frac{\sinh \alpha \ell^*\; \sinh \alpha (L-\ell^*)}{\alpha \ell^*\; \sinh \alpha L}
\right|
\right)^2
\right].\hspace*{6mm}
\label{smallf}
\end{eqnarray}
Integrations in this equation can be carried out analytically, resulting in a long expression that we do not reproduce here. Then the integral in Eq.\ (\ref{ftildeomega}) can be calculated numerically.

\begin{figure}[t]
\hspace*{-1.5cm}
\includegraphics[width=2.0\columnwidth]{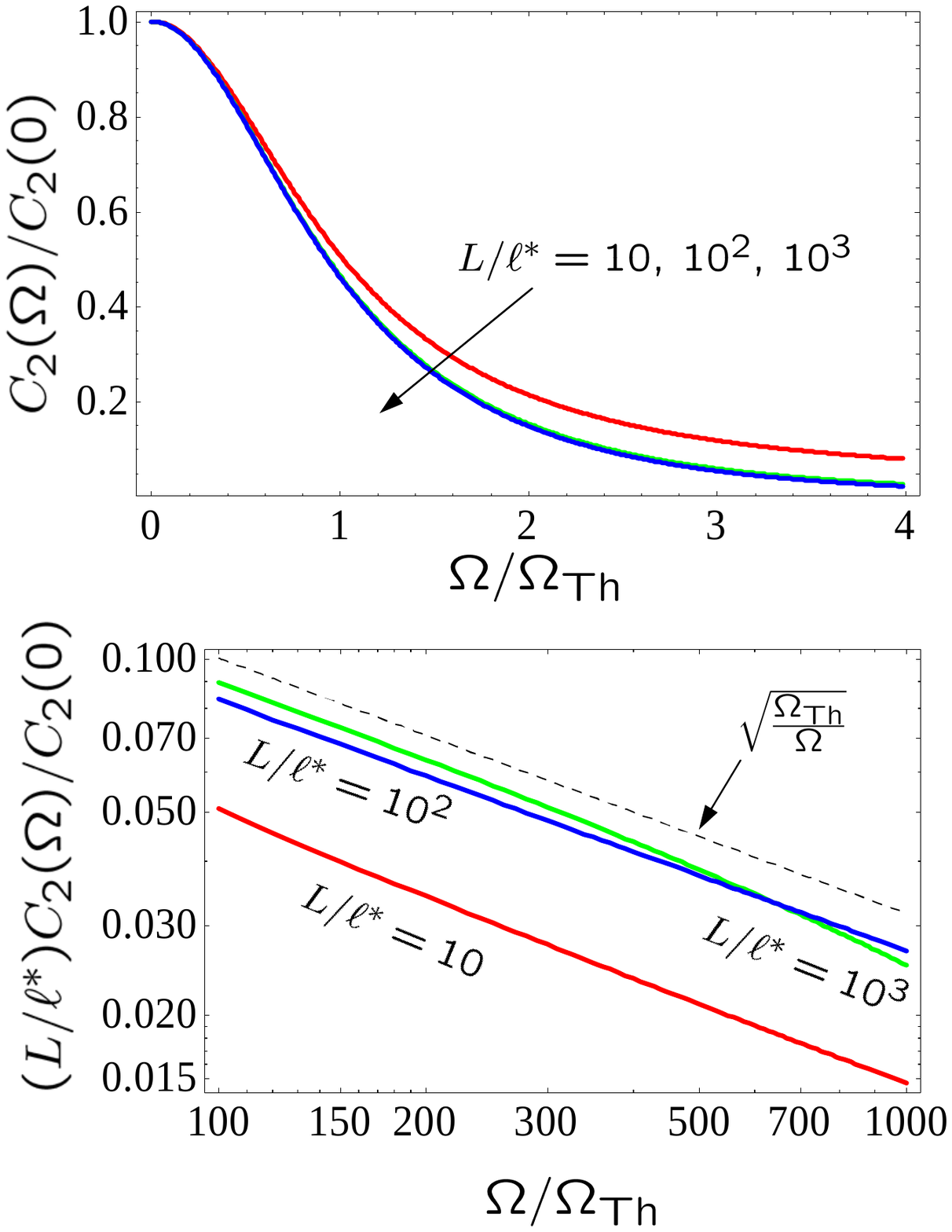}
\vspace*{-1.5cm}
\caption{\label{figc2slab}
Frequency correlation of intensity fluctuations $C_2$ in transmission of a plane wave through a disordered slab for small (top) and large (bottom) values of $\Omega$.}
\end{figure}

In the limit of a thick slab $L \gg \ell^*$, Eq.\ (\ref{c2slabomega}) yields a function that depends mainly on $\Omega/\Omega_{\mathrm{Th}}$ as far as $\Omega/\Omega_{\mathrm{Th}} \lesssim 1$, see Fig.\ \ref{figc2slab} (top). In the limit of large  $\Omega/\Omega_{\mathrm{Th}} \to \infty$ we find
\begin{eqnarray}
C_2^{\mathrm{plane\; wave}}(\Omega) \propto
C_2^{\mathrm{plane\; wave}}(0) \times
\frac{\ell^*}{L} \sqrt{\frac{\Omega_{\mathrm{Th}}}{\Omega}},
\label{largeomega}
\end{eqnarray}
as illustrated in Fig.\ \ref{figc2slab} (bottom).

\begin{figure}[t]
\hspace*{-1.5cm}
\includegraphics[width=2.0\columnwidth]{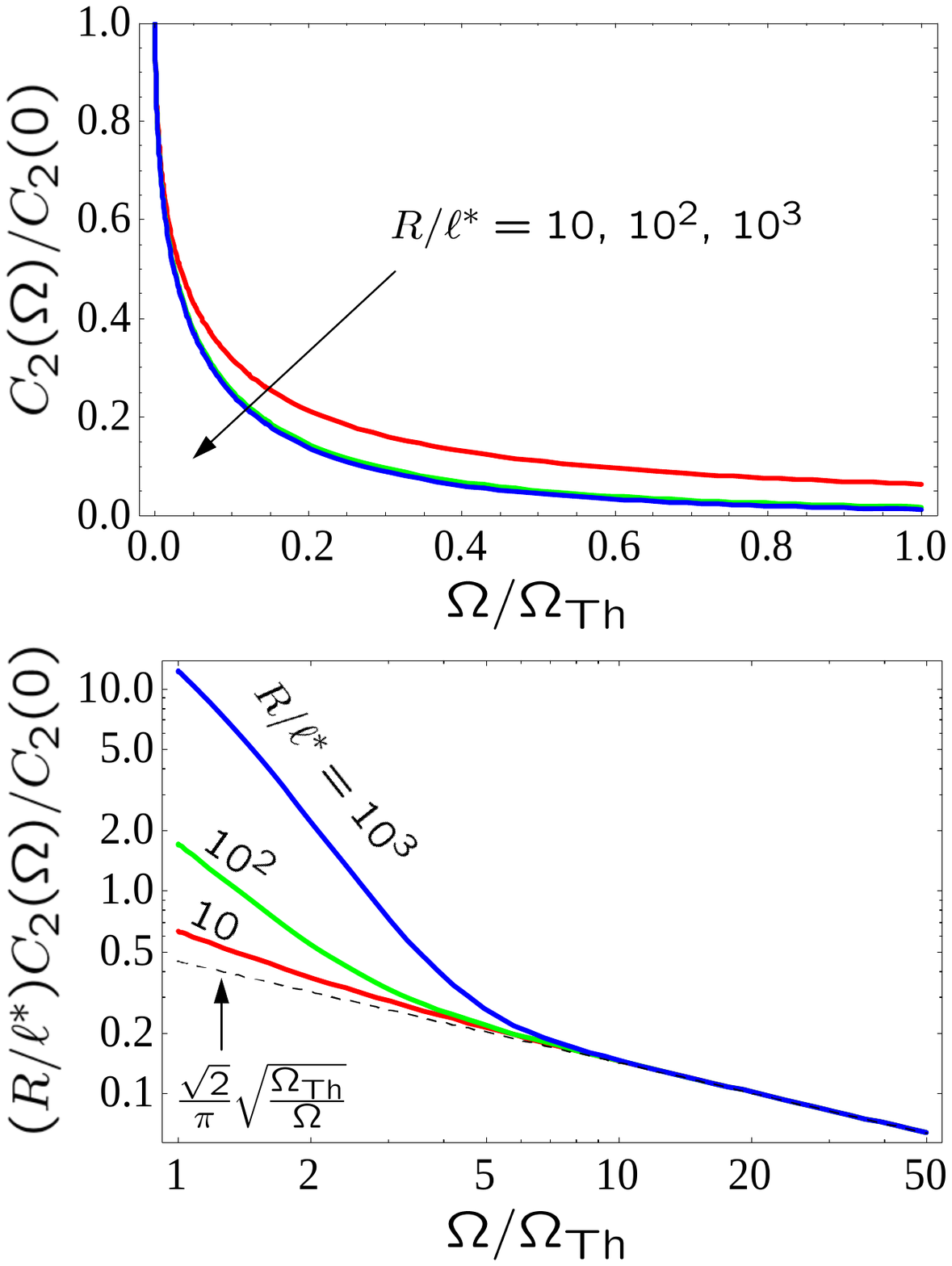}
\vspace*{-1.5cm}
\caption{\label{figc2inf}
Frequency correlation of intensity fluctuations $C_2$ at a distance $R$ from a point source in the infinite medium for small (top) and large (bottom) values of $\Omega$.}
\end{figure}

{\bf Point source in the infinite medium.}
Here we need to introduce a spatial cut-off $\sim \ell^*$ to avoid the path crossing (Hikami box) being closer than $\ell^*$ to the detector. The results then should be understood as depending on the precise value of this cutoff:
\begin{eqnarray}
C_2^{\mathrm{point\; source}}(\Omega) \simeq
\frac{3}{4(k \ell^*)^2} F_2 \left( \alpha R, \frac{\ell^*}{R} \right),
\label{c2infomega}
\end{eqnarray}
where
\begin{eqnarray}
F_2 \left( \alpha R, \frac{\ell^*}{R} \right) &=&
2 \frac{\ell^*}{R}
\left\{
\int\limits_0^{1-\ell^*/R} dx
\frac{\exp[-2 \mathrm{Re}(\alpha R) x]}{(1-x^2)^2}
\right.
\nonumber \\
&+& \left.
\int\limits_{1+\ell^*/R}^{\infty} dx
\frac{\exp[-2 \mathrm{Re}(\alpha R) x]}{(1-x^2)^2}
\right\}.
\label{ftildeinfomega}
\end{eqnarray}
In the limit of $R \gg \ell^*$ that is of interest for us here, we have
\begin{eqnarray}
F_2 \left( 0, \frac{\ell^*}{R} \right) &=& 1,
\\
F_2 \left( \alpha R, \frac{\ell^*}{R} \right) &=& \frac{\ell^*}{R} \times
\frac{1}{\mathrm{Re}(\alpha R)},\;\;\;
\Omega \gg \Omega_{\mathrm{Th}}
\label{flimit2}
\end{eqnarray}
The behavior of $C_2$ at small and large $\Omega$ is illustrated in Fig.\ \ref{figc2inf}.

{\bf Transmission of a tightly focused beam through a slab.}
We assume that the $C_2$ correlation function for a beam focused on the surface of a disordered slab is similar to the one for the plane wave, except for the magnitude of $C_2$ \cite{deboer92}:
\begin{eqnarray}
C_2^{\mathrm{focused\; beam}}(\Omega) \simeq
\frac{\mathrm{const}}{(k \ell^*)^2}
F_2\left(\alpha L, \frac{\ell^*}{L} \right),
\label{c2focusomega}
\end{eqnarray}
with $F_2(\alpha L, \ell^*/L)$ defined by Eq.\ (\ref{ftildeomega}).

\subsection{Infinite-range correlation $C_0$}

\begin{figure}[t]
\hspace*{-1.6cm}
\includegraphics[width=1.5\columnwidth]{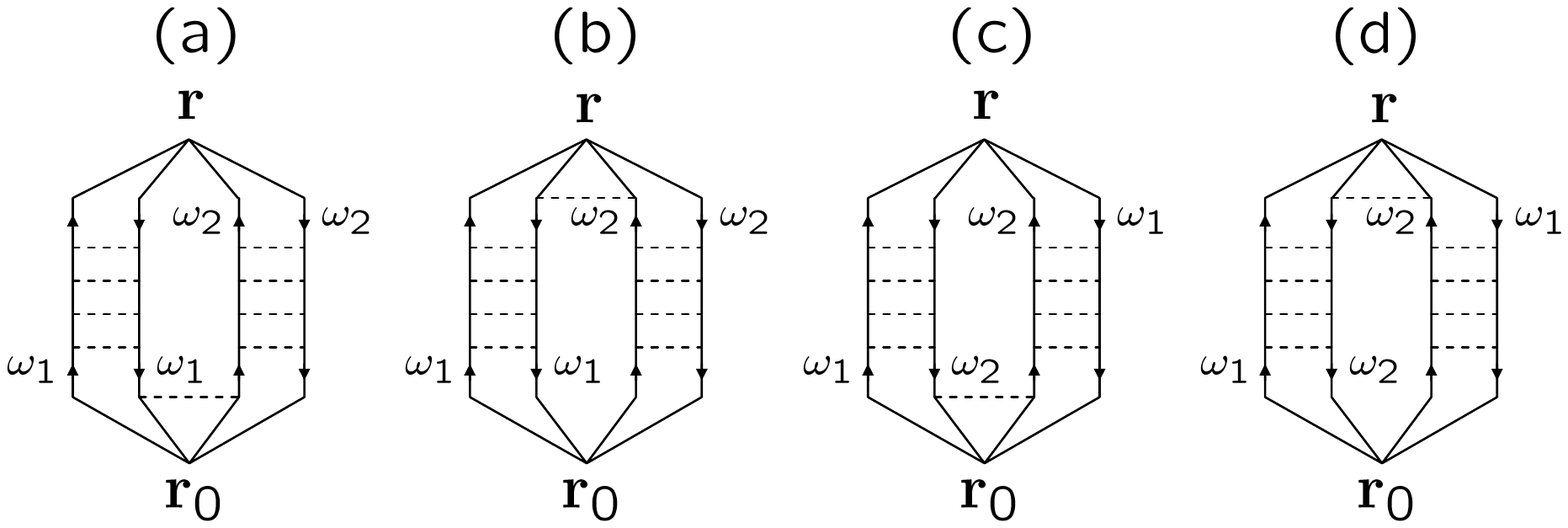}
\vspace*{-6cm}
\caption{Diagrams contributing to $C_0$ correlation function. $\vec{r}_0$ is the source position; $\omega_{1,2} = \omega \pm \frac12 \Omega$. The diagrams (a) and (b) are independent of $\Omega = \omega_1 - \omega_2$ as far as $|\Omega| \ll \omega_{1,2}$. A complex conjugate diagram should be added to each of the diagrams.}
\label{diagramsc0omega}
\end{figure}

The frequency dependence of the $C_0$ correlation function is obtained by calculating the diagrams of Fig.\ \ref{diagramsc0omega}. We obtain:
\begin{eqnarray}
C_0^{(a)}(\Omega) &=&
C_0^{\mathrm{(in)}},
\label{c0aomega}
\\
C_0^{(b)}(\Omega) &=&
C_0^{\mathrm{(out)}},
\label{c0bomega}
\\
C_0^{(c)}(\Omega) &=&
C_0^{\mathrm{(in)}} C_1(\Omega),
\label{c0comega}
\\
C_0^{(d)}(\Omega) &=&
C_0^{\mathrm{(out)}} C_1(\Omega).
\label{c0domega}
\end{eqnarray}

\subsection{Full expression for the correlation function}

Adding up all the contributions and assuming (as in the case of spatial correlations) that the behavior of $C_3$ as a function of frequency is similar to that of $C_2$, we finally find the full expression for the frequency correlation function:
\begin{eqnarray}
C(\Omega) &=& \left[1 + C_0^{\mathrm{(in)}} + C_0^{\mathrm{(out)}} \right] C_1(\Omega)
\nonumber \\
&+&
2A \times \frac{F_2(\alpha L, \ell^*/L)}{F_2(0, \ell^*/L)}
+ C_0^{\mathrm{(in)}} + C_0^{\mathrm{(out)}},\;\;\;
\label{fullomega}
\end{eqnarray}
where $A \sim 1/(k\ell^*)^2$.
Note that the constants $A$ and $C_0^{\mathrm{(in, out)}}$ here should be the same as in Eq.\ (\ref{cfull}) for the spatial correlation.


\begin{thebibliography}{29}

\bibitem{Anderson1958}
P.\,W. Anderson,
Phys. Rev. \textbf{109}, 1492 
(1958).

\bibitem{Abrahams1979a}
E. Abrahams, P.\,W. Anderson, D.\,C. Licciardello, and T.\,V. Ramakrishnan,
Phys. Rev. Lett. \textbf{42}, 673 
(1979).

\bibitem{Vollhardt1980b}
D. Vollhardt and P. W\"{o}lfle,
Phys. Rev. Lett. \textbf{45}, 842
(1980).

\bibitem{Evers2008a}
F. Evers and A.\,D. Mirlin,
Rev. Mod. Phys. \textbf{80}, 1355 (2008).

\bibitem{Sheng2006a}
P. Sheng,
\textit{Introduction to Wave Scattering, Localization and Mesoscopic Phenomena}
(Springer, Berlin, 
2006).

\bibitem{Abrahams2010}
E. Abrahams, ed.,
\textit{50 Years of Anderson Localization}
(World Scientific, Singapore, 2010).


\bibitem{John1983}
S. John, H. Sompolinsky, and M.\,J. Stephen,
Phys. Rev. B \textbf{27} 5592 (1983).

\bibitem{John1984}
S. John,
Phys. Rev. Lett. \textbf{53}, 2169 
(1984).

\bibitem{Anderson1985}
P.\,W. Anderson,
Phil. Mag. B \textbf{52}, 505 
(1985).

\bibitem{Hu2008b}
H. Hu, A. Strybulevych, J.\,H. Page, S.\,E. Skipetrov, and B.\,A. van Tiggelen,
Nature Phys. \textbf{4}, 945 (2008).
	
\bibitem{Sperling2013}
T. Sperling, W.\,B\"{u}hrer, C.\,M. Aegerter, and G. Maret,
Nature Photon. \textbf{7}, 48 
(2013).

\bibitem{Feng1988}
S. Feng, C. Kane, P.\,A. Lee and A.\,D. Stone,
Phys. Rev. Lett. \textbf{61}, 834 (1988).
	
\bibitem{DeBoer1992}
J.\,F. de~Boer, M.\,P. van Albada, and A. Lagendijk,
Phys. Rev. B \textbf{45}, 658 (1992).

\bibitem{Berkovits1994}
R. Berkovits and S. Feng,
Phys. Rep. \textbf{238}, 135 (1994).

\bibitem{Scheffold1997a}
F. Scheffold, W.\,H\"{a}rtl, G. Maret, and E. Matijevi\'{c},
Phys. Rev. B \textbf{56}, 10942 (1997).

\bibitem{VanRossum1999a}
M.\,C.\,W. van Rossum and T.\,M. Nieuwenhuizen,
Rev. Mod. Phys. \textbf{71}, 313 (1999).

\bibitem{Sebbah2000}
P. Sebbah, R. Pnini, and A.\,Z. Genack,
Phys. Rev. E \textbf{62}, 7348 (2000).

\bibitem{Chabanov2004}
A.\,A. Chabanov, N.\,P. Tr\'{e}gour\`{e}s, B.\,A. van~Tiggelen, and A.\,Z. Genack,
Phys. Rev. Lett.  \textbf{92}, 173901 (2004).

\bibitem{Akkermans2007}
E. Akkermans and G. Montambaux,
\textit{Mesoscopic Physics of Electrons and Photons} (Cambridge University Press, Cambridge, 2007).

\bibitem{Genack1999}
A. Z. Genack, P. Sebbah, M. Stoytchev, and B. A. van Tiggelen,
Phys. Rev. Lett. \textbf{82}, 715 (1999).

\bibitem{Cowan2007a}
M.\,L. Cowan, D. Anache-M\'{e}nier, W.\,K. Hildebrand, J.\,H. Page, and B.\,A. van~Tiggelen,
Phys. Rev. Lett. \textbf{99}, 094301 (2007).

\bibitem{Shapiro1999a}
B. Shapiro,
Phys. Rev. Lett. \textbf{83}, 4733 (1999).

\bibitem{Skipetrov2000}
S.\,E. Skipetrov and R. Maynard,
Phys. Rev. B \textbf{62}, 886 (2000).

\bibitem{VanTiggelen2006b}
B.\,A. van Tiggelen and S.\,E. Skipetrov,
Phys. Rev. E \textbf{73}, 045601 (2006).

\bibitem{Caze2010}
A. Caz\'{e}, R. Pierrat and R. Carminati, Phys. Rev. A \textbf{82}, 043823 (2010).

\bibitem{Birowosuto2010a}
M.\,D. Birowosuto, S.\,E. Skipetrov, W.\,L. Vos, and A.\,P. Mosk,
Phys. Rev. Lett. \textbf{105}, 013904 (2010).

\bibitem{Krachmalnicoff2010a}
V. Krachmalnicoff, E. Castani\'{e}, Y. De Wilde, and R. Carminati,
Phys. Rev. Lett. \textbf{105}, 183901 (2010).

\bibitem{Sapienza2011}
R. Sapienza, P. Bondareff, R. Pierrat, B. Habert, R. Carminati, and N.\,F. van Hulst,
Phys. Rev. Lett. \textbf{106}, 163902 (2011).

\bibitem{Garcia2012}
P.\,D. Garc\'{\i}a, S. Stobbe, I. S\"{o}llner, and P. Lodahl,
Phys. Rev. Lett. \textbf{109}, 253902 (2012).

\bibitem{El-Dardiry2011}
R.\,G.\,S. El-Dardiry, S. Faez, and A. Lagendijk,
Phys. Rev. A \textbf{83}, 031801 (2011).

\bibitem{Mirlin2000a}
A.\,D. Mirlin,
Phys. Rep. \textbf{326}, 259 (2000).

\bibitem{Dobrosavljevic2003a}
V. Dobrosavljevi\'{c}, A.\,A. Pastor, and B.\,K. Nikoli\'{c},
Europhys. Lett. \textbf{62}, 76 (2003).

\bibitem{Murphy2011}
N.\,C. Murphy, R. Wortis, and W.\,A. Atkinson,
Phys. Rev. B \textbf{83}, 184206 (2011).

\bibitem{Cinf}
  The possible existence of a small infinite-range
  contribution to 
  the intensity correlations was pointed out in
  Ref.\ \cite {Chabanov2004} for waves in quasi-1D disordered waveguides, but
  their origin was not identified with certainty.

\bibitem[{suppmat()}]{suppmat}
  See the supplemental material at the end of this document.

\bibitem{freqfit}
  The fits of the frequency correlations are performed
  similarly to those of the spatial correlations, with the additional parameter
  $\Omega _{\protect \mathrm {Th}}$ also constrained to have the same value for
  both single- and scanned-source experiments.

\bibitem{C0robust}
  Note that when both spatial and frequency correlations
  are measured, the values of both $C_0^{\protect \rm {(in)}}$ and
  $C_0^{\protect \rm {(out)}}$ can be determined most robustly from the
  asymptotic values of the correlations. For example, the fitted values of
  $C_0^{\protect \rm {(out)}}$ measured from the frequency correlations, which
  contain an infinite-range contribution due to $C_0^{\protect \rm {(out)}}$,
  are more reliable than those measured from the spatial correlations.

\bibitem{Turner1998a}
J.\,A. Turner, M.\,E. Chambers, and R.\,L. Weaver,
Acustica \textbf{84}, 628 (1998).

\bibitem{Yang2002}
S. Yang, J.\,H. Page, Z. Liu, M.\,L. Cowan, C.\,T. Chan, and P. Sheng,
Phys. Rev. Lett. \textbf{88}, 104301 (2002).

\bibitem{John1987}
S. John,
Phys. Rev. Lett. \textbf{58}, 2486 (1987).

\bibitem{Faez2009a}
S. Faez, A. Strybulevych, J.\,H. Page, A. Lagendijk, and B.\,A. van Tiggelen,
Phys. Rev. Lett.  \textbf{103}, 155703 (2009).

\end{thebibliography}

\begin{thebibliography}{11}


\bibitem{natphysloc}
H. Hu, A. Strybulevych, J.\,H. Page, S.\,E. Skipetrov, and B.\,A. van Tiggelen,
Nature Phys. \textbf{4}, 945 (2008).

\bibitem{page1996}
J.\,H. Page, P. Sheng, H.\,P. Schriemer, I. Jones, X. Jing, and D.\,A. Weitz,
Science \textbf{271}, 634 (1996).

\bibitem{classsource}
R.\,G.\,S. El-Dardiry, S. Faez, and A. Lagendijk,
Phys. Rev. A \textbf{83}, 031801 (2011).



\bibitem{akker07}
E. Akkermans and G. Montambaux,
\textit{Mesoscopic Physics of Electrons and Photons} (Cambridge University Press, Cambridge, 2007).

\bibitem{shapiro99}
B. Shapiro,
Phys. Rev. Lett. \textbf{83}, 4733 (1999).

\bibitem{shapiro86}
B. Shapiro,
Phys. Rev. Lett. \textbf{57}, 2168 (1986).

\bibitem{freund92}
I. Freund and D. Eliyahu,
Phys. Rev. A \textbf{45}, 6133 (1992).

\bibitem{sebbah00}
P. Sebbah, R. Pnini and A.\,Z. Genack,
Phys. Rev. E \textbf{62}, 7348 (2000).


\bibitem{carminati2010}
R. Carminati, Phys. Rev. A \textbf{81}, 053804, (2010).




\bibitem{stephen87}
M.\,J. Stephen and G. Cwilich,
Phys. Rev. Lett. \textbf{59}, 285 (1987).

\bibitem{pnini89}
R. Pnini and B. Shapiro,
Phys. Rev. B \textbf{39}, 6986 (1989).

\bibitem{retzker02}
A. Retzker and B. Shapiro,
Pramana \textbf{58}, 225 (2002).

\bibitem{vantiggelen06}
B.\,A. van Tiggelen and S.\,E. Skipetrov,
Phys. Rev. E \textbf{73}, 045601 (2006).

\bibitem{skip00}
S.\,E. Skipetrov and R. Maynard,
Phys. Rev. B \textbf{62}, 886 (2000).

\bibitem{deboer92}
J.\,F. de Boer, M.\,P. van Albada, and A. Lagendijk,
Phys. Rev. B \textbf{45}, 658 (1992)

\end{thebibliography}
\end{document}